\documentclass[12pt]{article}

\usepackage{graphicx}
\begin{document}

\begin{center}

{\bf \Large OPTICAL AND HI PROPERTIES OF ISOLATED GALAXIES IN THE 2MIG CATALOG. I. GENERAL RELATIONSHIPS}

{\bf Yu. N. Kudrya$^1$, V. E. Karachentseva$^2$, and I. D. Karachentsev$^3$}

$^1$ Astronomical Observatory, Taras Shevchenko Kiev National University, Ukraine; e-mail: yukudrya@ukr.net \\
$^2$ Main Astronomical Observatory, National Academy of Sciences of Ukraine, Ukraine; e-mail: valkarach@gmail.com\\
$^3$ Special Astrophysical Observatory, Russian Academy of Sciences, Russia; e-mail: ikar@sao.ru

\bigskip

{\bf \large Abstract}
\end{center}
\bigskip

We analyze empirical relationships between  the optical, near infrared, and HI characteristics of isolated galaxies from the 2MIG Catalog covering the entire sky. Data on morphological types, $K_S$-, and $B$-magnitudes, linear diameters, HI masses, and rotational velocities are examined. The regression parameters, dispersions, and correlation coefficients are calculated for pairs of these characteristics. The resulting relationships can be used to test the hierarchical theory of galaxy formation through numerous mergers of cold dark matter. 


\bigskip

{\bf \large 1. Introduction}

\bigskip

 According to modern ideas, the morphology, star formation processes, chemical composition, and other characteristics of galaxies depend strongly  on their surroundings. Isolated galaxies, which occur in regions of very low densities of matter, have not been subjected to significant interactions with their surroundings over the last few billion years. Thus, the population of isolated galaxies can serve as a reference sample for study of the origin and evolution of galaxies. This idea is widely accepted and the need to create a complete, homogeneous sample of these galaxies is obvious. In practice, however, we are dealing with the rather complicated problem of creating catalogs (lists) of isolated galaxies. Various way of detecting isolated galaxies in 2D and 3D spaces have been developed [1]. These have been
applied to different surveys of the sky in various wavelengths [2--4]. The methods for selecting isolated galaxies often are applied to small segments of the sky or to different volumes. Finally, studies have been made of specific samples of isolated galaxies established by morphology [5,6], the existence of active nuclei [7], etc. The various approaches to studying isolated galaxies and results from these studies were presented at the conference "Galaxies in Isolation: Nature vs. Nurture" [1].

 Up to now, the most fully studied catalog has been the Catalog of Isolated Galaxies (CIG) [8]. The following empirically chosen relations were used as criteria for compiling this catalog: 
$$x_{1i} \geq 20a_i, \eqno(1) $$

$$ 1/4 a_1\leq  a_i\leq  4a_1, \eqno(2)$$
where the subscripts "1" and "$i$" refer, respectively, to the detected galaxy and to neighboring galaxies. In other words, a galaxy with an angular diameter $a_1$ is considered to be isolated if all the "significant" neighbors with angular diameters $a_i$ are located at distances $x_{1i}$ from it no closer than $20a_i$. After visual examination of the O and E prints of the Palomar Sky Survey POSS--1 of all the galaxies from the Zwicky catalog CGCG [9] and their neighborhoods, 1051 galaxies met the isolation criterion  ($m\leq15.7^m,  \delta\geq-3^{\circ},\mid b\mid\geq 20^{\circ}$); these are roughly 3\%
 of the entire number of galaxies in the CGCG.

 Since the beginning of the 21-st century an international team (Spain, USA, France, Italy) has been engaged in an extensive project, AMIGA (http://www.iaa.es/AMIGA.html), for studying the physical properties of the most isolated galaxies of the CIG catalog and the interstellar medium within them, taking new observational data into account. In the course of this work, the high efficiency of the isolation criterion  has been tested and confirmed for most of the galaxies of the CIG catalog. 

A new catalog of isolated galaxies, 2MIG [10] (electronic version at ftp://cdsarc.u-strasb.fr/pub/cats/\\YII/257), has recently been published. It was created using the advantages of the photometrically uniform Two Micron All-Sky Survey (2MASS) [11] which covers the entire sky. The 2MIG catalog is based on automatic sampling of extended sources from the 2MASS Extended Source Catalog (2MASS XSC) [12] with subsequent visual review. A modification of conditions (1) and (2) for the objects in the 2MASS XSC catalog was used to identify the isolated 2MIG galaxies. The dimensionless "distance" $x_{1i}/a_i\equiv x_{1i}/2r_i\equiv s_i$ under condition (1) was taken to be $s = 30$ since the infrared diameters of the galaxies in 2MASS are systematically smaller than their standard optical diameters $a_{25}$. A galaxy "1" with a $K$ magnitude $K_{20fe}$ and an isophotal $K$ diameter $a_K=2 r_{20fe}$ was considered to be isolated if conditions (1) and (2) are satisfied for it and a neighboring "significant" galaxy with $s = 30$. In order to have a substantial inventory of faint neighboring galaxies, in creating the 2MIG catalog the algorithm for identifying isolated galaxies was applied to all candidate galaxies with apparent magnitudes within the range

$$4.0^m<K\leq 12.0^m \eqno(3)$$
and angular diameters 

$$a_K\geq30^{\prime\prime}. \eqno(4)$$

The limiting apparent magnitude $K_S = 12.0^m$ was chosen so that it corresponded to the limit in the CIG catalog
for galaxies with typical color indices $B-K_S=3.5^m\div4.0^m$. The limitation on the stellar magnitude at the bright
end is a consequence of features of the photometry of the most extended bright galaxies in 2MASS. Since the 2MASS XSC catalog did not include objects with diameters $a_K<10^{\prime\prime}$, in order to satisfy condition (2) it was necessary to restrict both the candidate and isolated galaxies to a limiting diameter. The minimum angular size was taken to be 30$^{\prime\prime}$ for them.

 When the modified condition (1) and the conditions (2)--(4) were used, a total of 4025 candidate isolated galaxies were detected in the 2MASS XSC catalog; this represents 4045/51572 = 7.8\%
 of the galaxies brighter than $K_S = 12.0^m$ and larger than $a_K=30^{\prime\prime}$. In order to account for neighboring galaxies with a low surface brightness that are not usually visible in the 2MASS survey, other sky surveys, DSS--1, DSS--2, and SDSS, were also used. An additional test of isolation was carried out using information on the radial velocities from the HyperLEDA and NED data bases for the candidate objects and their nearest neighbors. After elimination of the non-isolated galaxies, 3227 objects, or 6.2\%
 of the number of galaxies brighter than $K_S = 12.0^m$ with diameters greater than $a_K=30^{\prime\prime}$ in the 2MASS XCS catalog, were included in the 2MIG catalog. The procedure for identifying isolated galaxies and the general properties of the 2MIG catalog have been described in detail elsewhere [10]. 

In this paper we give a statistical description of the optical and HI properties of the galaxies in the 2MIG catalog that have been identified in the HyperLEDA database and derive relationships among the different characteristics of the galaxies as functions of their morphological type.

\bigskip
\newpage
{\bf \large 2. Distributions of the main observational characteristics}

\bigskip

 Of the 3227 galaxies in the 2MIG catalog, 3070 were identified with objects in the HyperLEDA database. Their distribution on the moprphological type (in numerical code) is shown   in Fig. 1. All elliptical galaxies are assigned a value of --2 in the catalog. 

This sample contains 18\%
 elliptical and lenticular galaxies, 42\%
 Sa--Sab--Sb spirals with a dominant bulge, 38\%
 Sbc-Sc--Scd--Sd spirals with a dominant disk, and only 2\%
 irregular galaxies. 

We compared the types of 1864 galaxies common to the catalog and the HyperLEDA database. The results of this comparison are shown in Fig. 2. The solid line is the diagonal and the dashed line is a regression fit of  Type(HyperLEDA)  1.024$\cdot$Type(2MIG)--0.41. It can be seen that, on the average, the type in HyperLEDA is somewhat lower than that in the 2MIG catalog (by the difference between the dashed and solid lines). The standard deviation from the regression fit is equal to 3.6. 3\%
 of the galaxies lie beyond this value, in roughly equal amounts on both sides. We re-examined all the galaxies for which the moduli of the code differences  equalled to or exceeded 5. In those cases where the types were estimated in 2MIG as earlier, an independent classification
confirmed the validity of the types in 2MIG for 26 out of 30 cases (87\%
). In the opposite situation, the estimated type in 2MIG was confirmed for 24 out of 32 cases (75\%
). Note that 22 of these 62 galaxies are in the strong Galactic
 absorption zone ($A^G_B>2$), where it is very difficult to determine the morphological type. Further on in this paper
we assume that the classification of the galaxies according to 2MIG is more accurate. 

\begin{figure}
\begin{center}
\includegraphics[width=.5\textwidth]{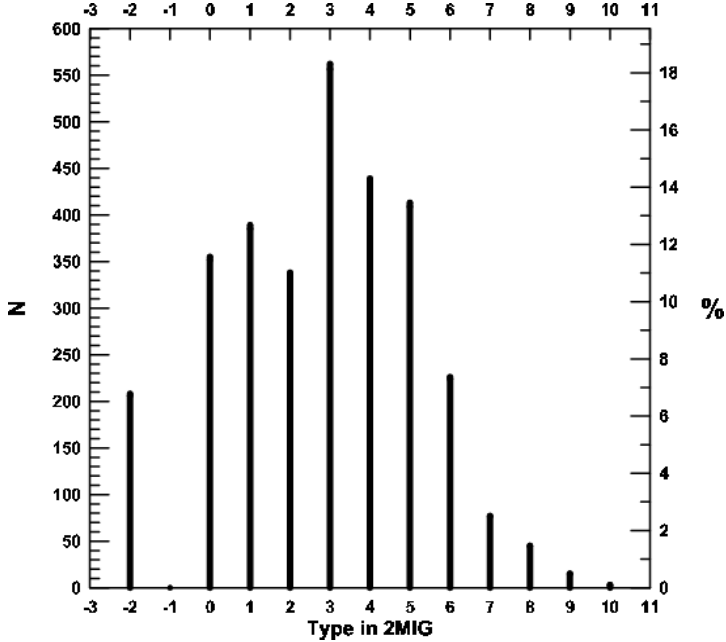}
\caption{Distribution of 3070 galaxies with respect to morphological type.}
\end{center}
\end{figure}

\begin{figure}
\begin{center}
\includegraphics[bb=93 128 508 596,clip,width=.5\textwidth]{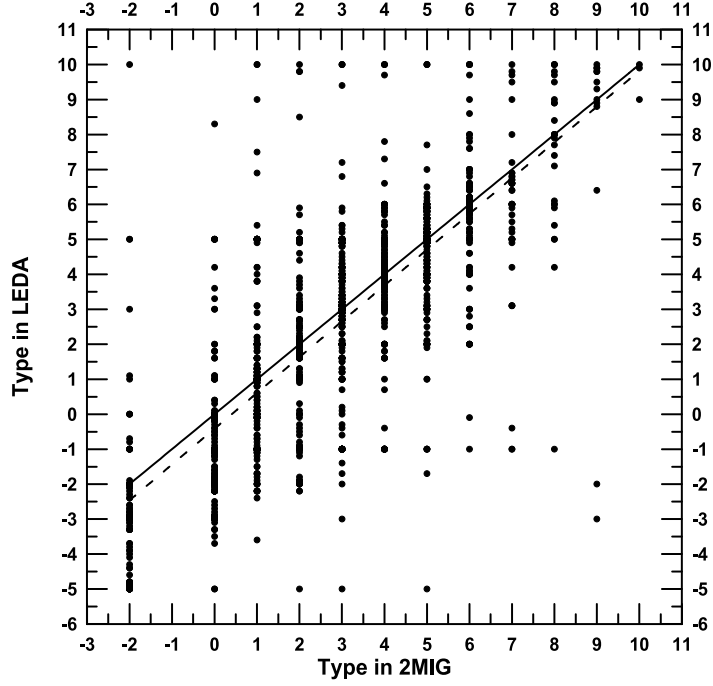}
\caption{Comparison between the morphological types of galaxies in the 2MIG catalog and HyperLEDA.}
\end{center}
\end{figure}

The inclination angles of the galaxies are given  for 2850 of the 2MIG galaxies in HyperLEDA. Of these, 40 have an inclination $i = 0^{\circ}$ and 450 are  visible practically edge-on from the edge ($i=85^{\circ}\div90^{\circ}$).

\begin{table}
\begin{center}
\caption{Parameters of the Distributions of the 2MIG Galaxies with Respect to the Main Characteristics}
\begin{tabular}{|l|c|c|c|r|r|r|} \hline\hline

n     &  Sample  & Characteristic& Mean &Min & Max& $\sigma$\\
\hline
1     &  N = 3227& $K_S$         & 10.94 &  5.04  &12.00 &0.814\\
2     &  N = 3070& $K_S^{corr}$       & 10.84 &  4.81  &11.99& 0.832\\
3     &  N = 2721& $B_t$            & 15.15 &  10.41 &19.05& 1.161\\
4     &  N = 2703& $B_t^{corr}$       & 14.26 &  8.30  &18.57& 1.138\\
5     &  N = 2852& $\log(a_{25})$    & 0.007 &  --0.99 &1.01 & 0.230\\
6     &  N = 1849& $\log(a_{25}^{corr})$  & 0.129 &  --0.72 &1.15 & 0.196\\
7     &  N = 2638& SB            & 23.67 &  20.58 &27.55& 0.636\\
8     &  N = 2510& $V_h$(2MIG)     & 6818  &  --100  &23773& 3628\\
9     &  N = 2494& $V_{LG}$           & 6758  &  --304  &23760& 3645\\
10    &  N = 1106& $V_{rot}$          & 175.6 &  12.68 &613.9& 69.5\\
11    &  N = 1106& $\log(2V_{rot})$    & 2.509 &  1.404 &3.089& 0.190\\
12    &  N = 814& $\log(2V_{rot})$     &2.509  & 1.404 &3.089 & 0.186\\
      & $i>50^{\circ}$&                      &       &       &      & \\
\hline
\end{tabular}
\end{center}
\end{table}
We found no significant dependence of the average angle of inclination on type, which indicates that there is no selection on  this parameter. 

The parameters of the distribution of the main observable characteristics of the 2MIG galaxies are given in Table 1. The $K_S$ magnitudes and radial velocities $V_h$ are taken from 2MIG and the other data, from HyperLEDA. The table lists average, minimum, and maximum values as well as the standard deviation from the mean. The first and second columns of Table 1 show the parameters of the distribution on  the uncorrected catalog magnitudes
corr $K_S$ for the complete set of $N = 3227$ galaxies and on the corrected magnitudes $K_S^{corr}$ for the set of $N = 3070$ galaxies identified in HyperLEDA. We have corrected the magnitudes for the Galactic absorption and for internal absorption: $K_S^{corr}=K_S-A^G_K - A_K^{in}$ . The values of $A_B^G$ and $A_B^{in}$ were taken from HyperLEDA for the $B$-band
and transformed for the $K_S$ magnitudes with a coefficient of 0.084 in accordance with  [13]. The internal absorption $A_B^{in}$  is given not for  all the 3070 galaxies in HyperLEDA; when this was lacking, it was assumed equal to zero. The maximum value $K_S = 12^m$ corresponds to the limit in the 2MIG catalog. A comparison of the characteristics of the distributions with respect to $K_S$ and $K_S^{corr}$
reveals only some small differences. In particular, the average correction 
is $0.1^m$. Histograms of the distribution of the 3227 galaxies  to $K_S$ and of the 3070 galaxies on
to $K_S^{corr}$ are shown in Fig. 3a.

\begin{figure}
\includegraphics[bb=87 200 508 596,clip,width=.35\textwidth]{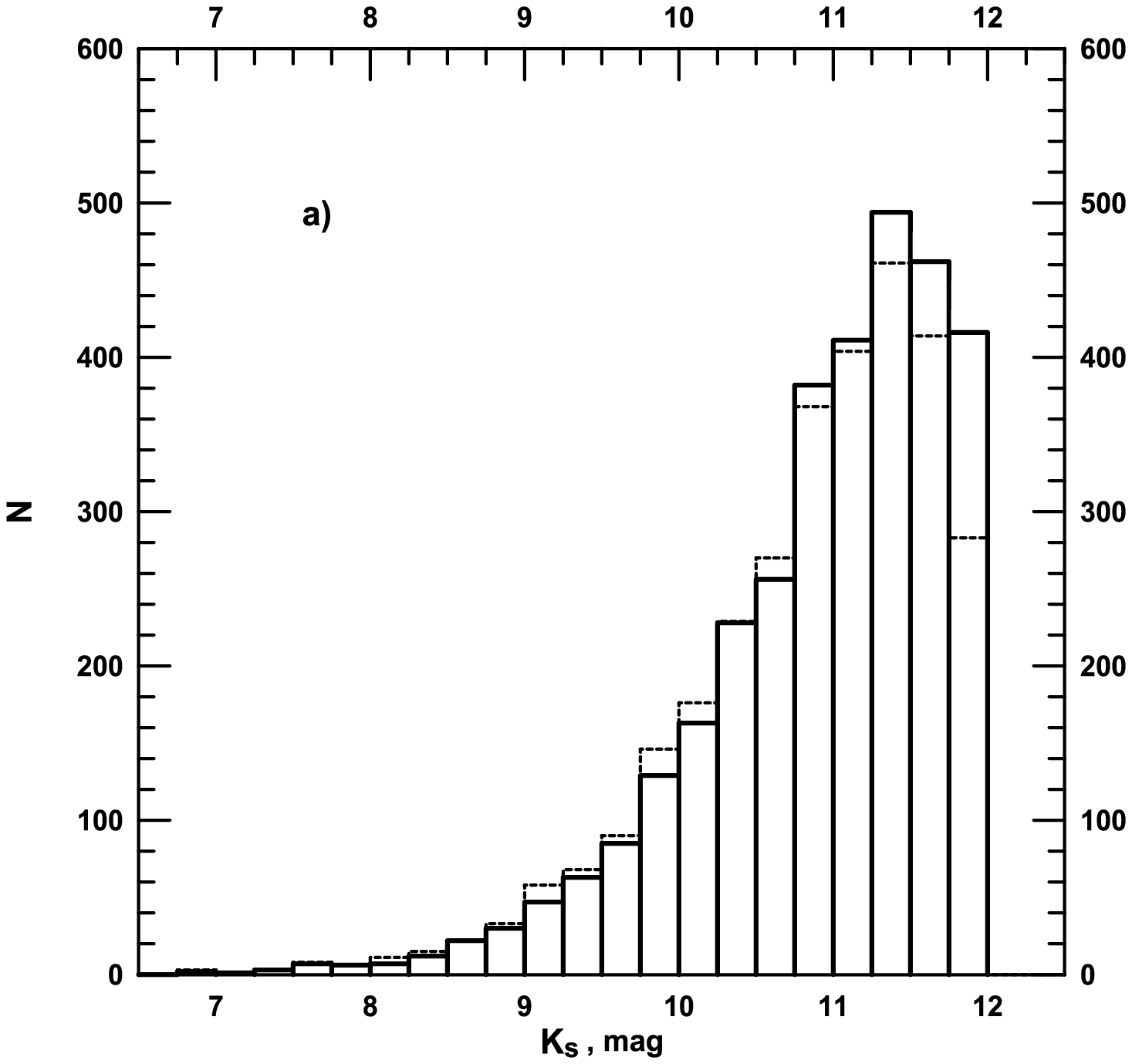}
\includegraphics[bb=87 198 508 596,clip,width=.35\textwidth]{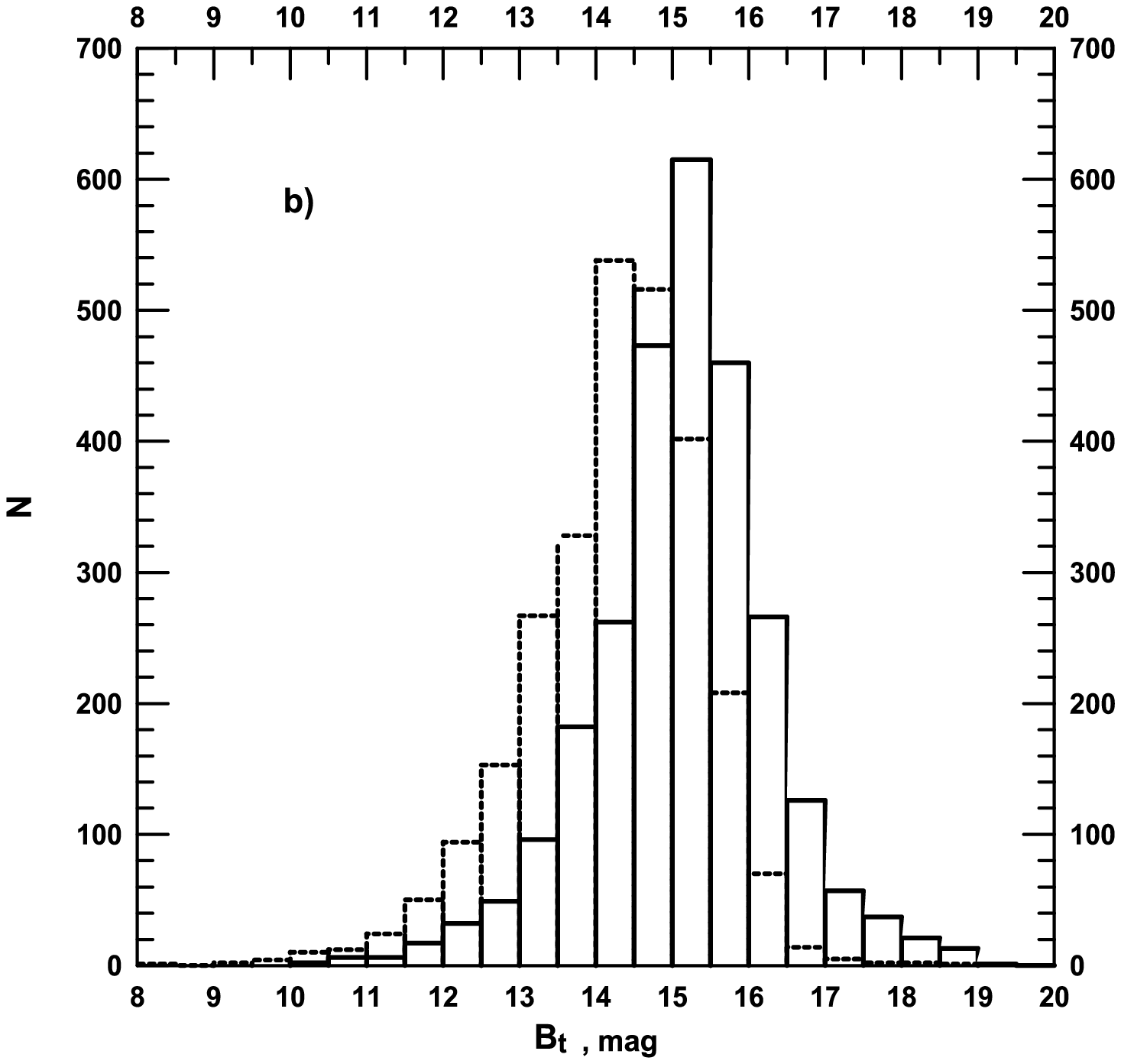}\\ \hfill
\includegraphics[bb=86 127 508 525,clip,width=.35\textwidth]{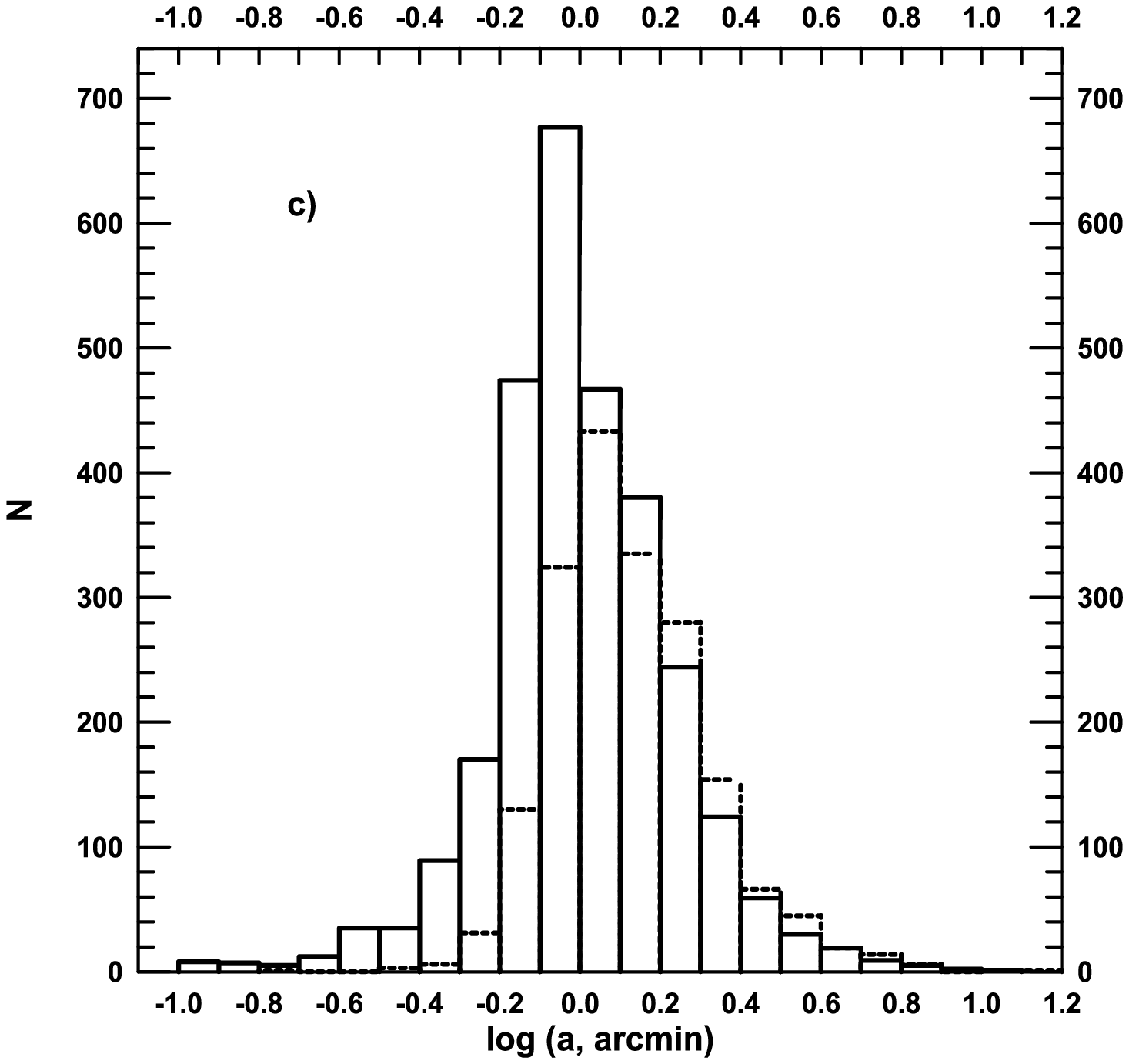}
\includegraphics[bb=86 200 509 596,clip,width=.35\textwidth]{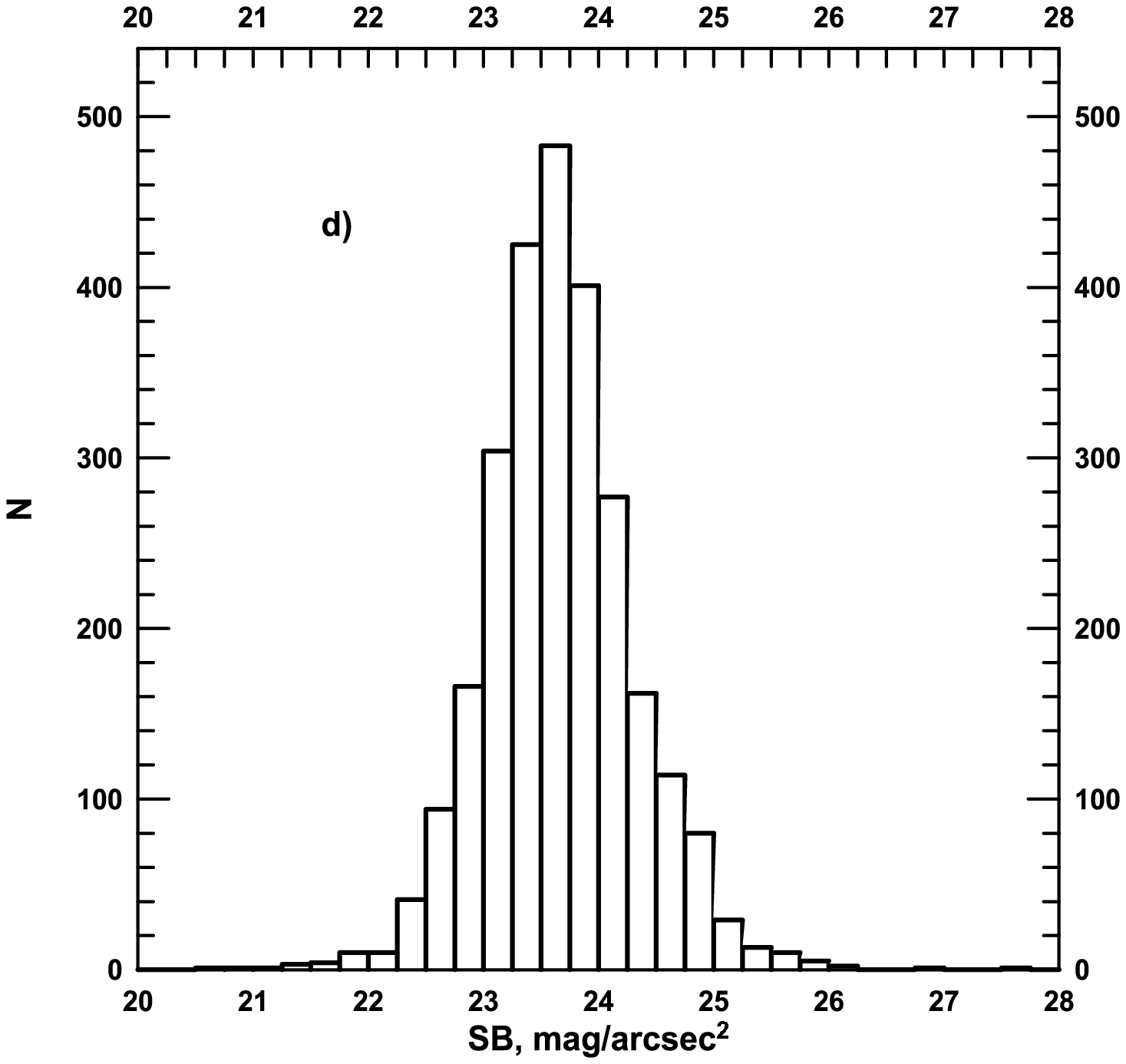} \hfill
\caption{Distributions of the 2MIG galaxies on: (a) $K_S$ (solid lines) and $K_S^{corr}$ (dashed lines); (b) $B_t$ (solid lines) and $B_t^{corr}$ (dashed lines); (c) $\log(a_{25}$) (solid lines) and $\log(a_{25}^{corr}$) (dashed lines); and (d) $B$ surface brightness within the 25 mag/arcsec$^2$ isophote.}
\end{figure}

The characteristics of the distributions on $B_t$ and $B_t^{corr}$ are shown in rows 3 and 4 of Table 1. It can be seen that they differ quite substantially. The difference, $0.89^m$, in the averages of $B_t$ and $B_t^{corr}$ is determined by the average value of the corrections and by the fact that the 18 galaxies without $B_t^{corr}$ in HyperLEDA are fairly faint, with an average magnitude $B_t = 17.65^m$. The distribution of the galaxies  on the magnitudes $B_t$ and
$B_t^{corr}$ is shown in Fig. 3b.

The parameters of the distributions of the 2852 2MIG galaxies on the logarithm of the uncorrected isophotal angular diameters up to an isophote of 25 mag/arcsec$^2, \log(a_{25}$), and of the 1849 galaxies on the logarithms of the angular diameters corrected for Galactic absorption and internal absorption, $\log(a_{25}^{corr}$ ), are given
in rows 5 and 6 of the Table 1 and are shown in Fig. 3c. 

The mean surface brightness within the 25 mag/arcsec$^2 B$ isophote (SB) was calculated from the $B$ magnitudes by averaging over a circle of diameter $a_{25}$ taking the Bottinelli correction described in HyperLEDA into account. The parameters of the distribution on SB are given in row 7 of Table 1. Note that 61 objects (2.3\%
) have SB$>25$ mag/arcsec$^2$ (in conflict with the definition of the average surface brightness in HyperLEDA). We assume that data for the magnitudes and diameters of the galaxies in HyperLEDA were determined independently from different sources and that, in some cases, a formal calculation of the mean surface brightness led to erroneous values. The distribution of the 2638 galaxies with respect to mean surface brightness is presented in Fig. 3d.

 Row 8 of Table 1 lists the parameters of the distribution of 2510 2MIG galaxies on their radial heliocentric velocities $V_h$ (known at the time the 2MIG catalog was published). The radial velocities $V_{LG}$ in the Local Group system were calculated from $V_h$ with the following parameters for the Sun's motion [14]: $V_{\odot}=  (316\pm5$) km/s,
 $l_0=93^{\circ}\pm2^{\circ}$, and $b_0=-4^{\circ}\pm2^{\circ}$. The parameters of the distribution of the 2494 galaxies on  $V_{LG}$ are listed in row 9 of Table 1. 

Figure 4a shows the distributions on the heliocentric radial velocity $V_h$ given in the catalog and the radial velocity $V_{LG}$  identified in HyperLEDA. The peak in the distribution for the 2MIG galaxies lies at $V_h \sim5000$ km/s. Note that two peaks were observed for the CIG galaxies [8], $V_h\sim1500$ km/s and $\sim6000$ km/s, which correspond to the galaxies of the Local supercluster and the Pisces-Perseus supercluster [15]. These differences are evidently affected by selectivity in choosing galaxies for inclusion in the catalog: isolated galaxies in the CIG were selected from an optical sample that included more blue nearby galaxies. However, the average depth of both catalogs is roughly the same at $\sim6500$ km/s. 

Rotation velocities $V_{rot}$ are presented in HyperLEDA for only 34\%
 of all the 3227 2MIG galaxies. The parameters of the $V_{rot}$ distribution  are given in row 10 of Table 1. Seven galaxies have $V_{rot} > 400$ km/s. While the high value of $V_{rot}$ for two of these can be explained by inaccuracies in the correction for the inclination (13$^{\circ}$ and 20$^{\circ}$), for the others there may be problems related to the observation conditions (low signal to noise ratio, etc.). Note that the HI observations by the ALFALFA group [16] also include galaxies with hydrogen line widths $W_{50}\approx  700-1000$ km/s. 

The distribution of the 1106 2MIG galaxies on the logarithm of twice the rotation velocity, $\log(2V_{rot}$), is shown in Fig. 4b (solid curve). Also shown in this figure is the distribution on 
$\log(2V_{rot}$) 
of the 814 galaxies for which the angle of inclination $i\geq  50^{\circ}$ (dashed curve). The characteristics of these two distributions are given in rows 11 and 12 of Table 1. They clearly differ little.

\bigskip
{\bf \large 3. The completeness of the 2MIG catalog}
\bigskip

 We studied the completeness of the 2MIG catalog using the "number of galaxies-- apparent magnitude" test for the uncorrected $K_S$ magnitudes assuming a uniform 3D distribution of the galaxies, $\log N(K_S) =  0.6 K_S +$ const, for all 3227 galaxies in the catalog. The resulting dependence is shown in Fig. 5 by the solid  curve. The dotted straight line has a slope of 0.6 and an arbitrary zero point. Clearly, the 2MIG sample is approximately complete over a range of five $K_S$ magnitudes, from $6.5^m$ to $11.5^m$.

We also checked the completeness of the sample of $N = 3070$ using the corrected magnitudes $K_S^{corr}$ . The
$N = 3070$ sample was complete over roughly the same limiting magnitudes. The bend in $\log N(K_S)$ at $K_S = 11^m- 12^m$ is caused both by the limits of the 2MIG catalog with respect to the $K_S$ magnitude and by the fact that the 2MASS XSC catalog did not include compact objects with diameters $a_K = 2r_{20 fe}<  10^{\prime\prime}$. 

For testing the uniformity and completeness of the catalog we have also used the Schmidt test [17]. For this, the average value of $V/V_{max}$ is calculated, where $V$ is the volume determined by the distance to a given object and $V_{max}$ is the maximum volume of the sample being considered. For a uniform distribution of galaxies in space without selection in the catalog, $\langle V/V_{max}\rangle=  0.5$, where the average is taken over the objects within a sample volume $V_{max}$. If $\langle V/V_{max}\rangle$ differs significantly from 0.5, then either the objects are clustered or selection is substantial.

As a crude approximation of the measure of the distance to an object, we use the  magnitude $K_S$ under
the assumption that all the objects have the same absolute luminosity. Then
\begin{figure}
\begin{center}
\includegraphics[bb=87 200 508 596,clip,width=.4\textwidth]{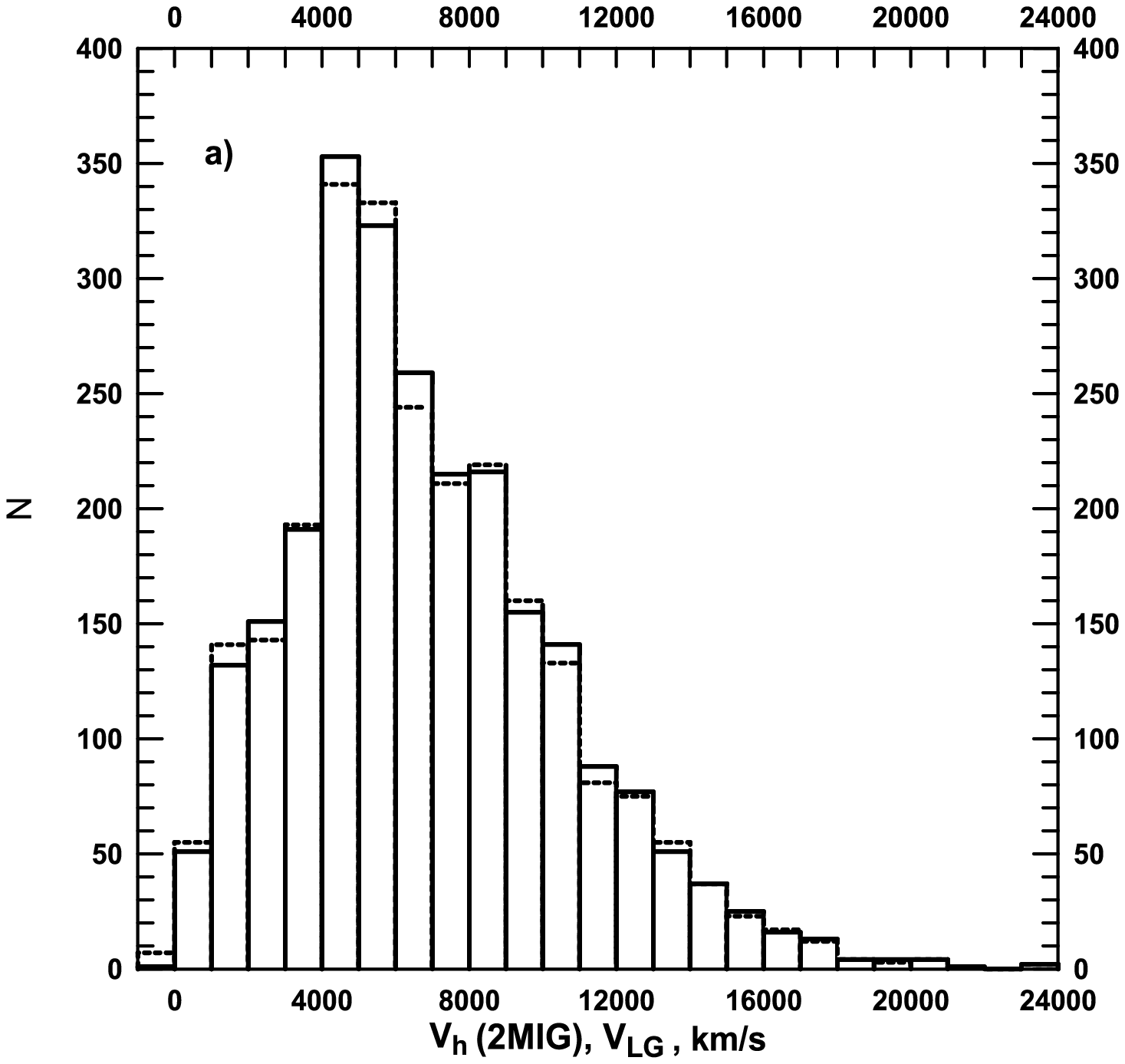}
\includegraphics[bb=87 198 508 596,clip,width=.4\textwidth]{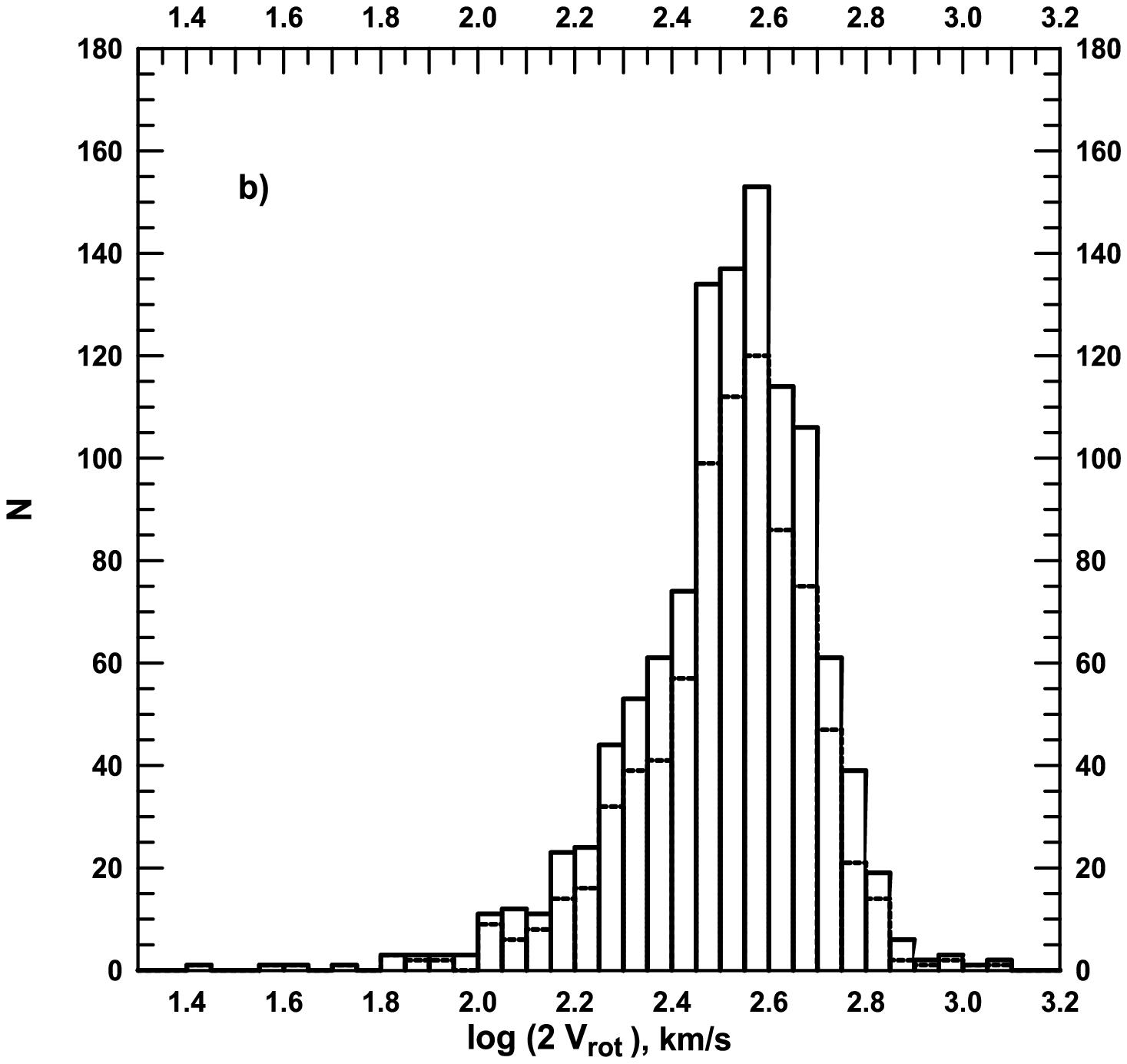}\\ \hfill
\caption{Distributions of the 2MIG galaxies on: (a) radial velocities $V_h$ (solid lines) and $V_{LG}$ (dashed lines) and (b) $\log(2V_{rot}$): all-- solid lines, galaxies with angles of inclination $i\geq 50^{\circ}$ -- dashed lines.} 
\end{center}
\end{figure}

\begin{figure}
\begin{center}
\includegraphics[bb=90 127 505 524,clip,width=.4\textwidth]{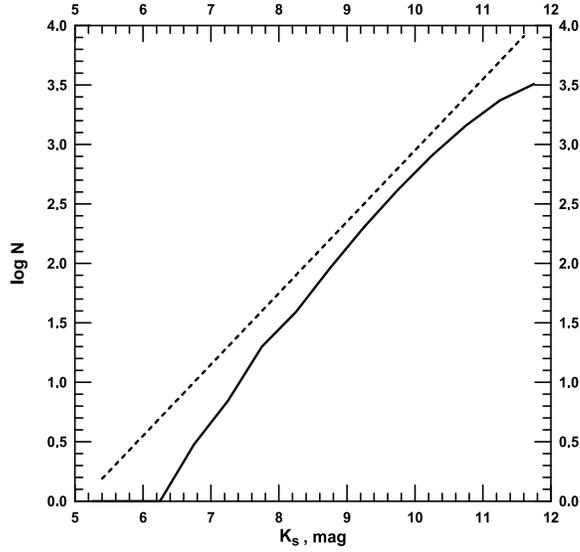}
\caption{The "logarithm of the number --stellar magnitude" dependence for galaxies of the 2MIG catalog as a function of the uncorrelated magnitude $K_S$.}
\end{center}
\end{figure}

\begin{figure}
\begin{center}
\includegraphics[bb=86 127 505 524,clip,width=.4\textwidth]{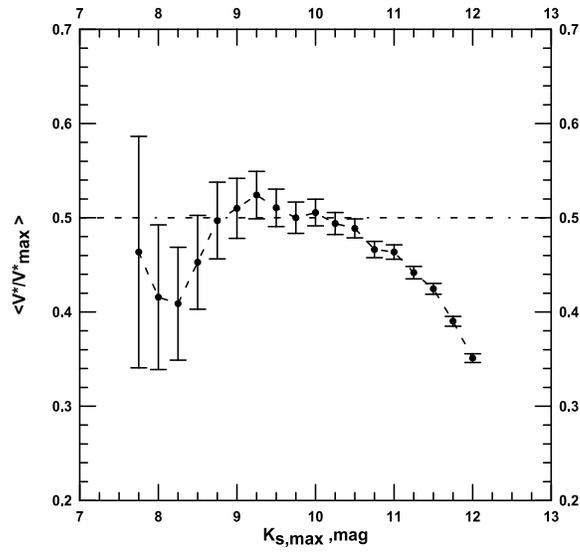}
\caption{Schmidt test for completeness and uniformity of the 2MIG catalog with respect to the uncorrected magnitude $K_S$.}
\end{center}
\end{figure}

$$V/V_{max}= (r/r_{max})^3  = dex[0.6\cdot (K_S - K_{S,max})]. \eqno(5)$$
  
To minimize the influence of the substantial inhomogeneity in the distribution of nearby objects, we exclude them, while starting to take the average at some minimum volume (denoted by $V^*=  V- V_{min}$ ), in our case starting at some minimum value $K_{S,min}$.

Figure 6 is a plot of $\langle V^*/V^*_{max}\rangle$ as a function of $K_{S,max}$ for all 3227 galaxies in the 2MIG catalog. (Here it was assumed that $K_{S,min}=  7.0^m$ ). Taking the completeness index $\langle V^*/V^*_{max}\rangle=  0.41$ [15], we find that the 2MIG catalog is roughly 80\%
 complete up to $K_S = 11.5^m$. This is roughly equivalent to the completeness of 80--90\% 
up to $m = 15.2^m$ in the CIG catalog of isolated galaxies [18].

\bigskip

{\bf \large  4. Distribution of basic absolute characteristics}
\bigskip

 To calculate the distribution of the absolute characteristics of the 2MIG galaxies we use an estimate of the distance $d$ in Mpc in accordance with the linear Hubble relation, $d = V_{LG}/H_0$ where $H_0=72$ km/s/Mpc. The rows in Table 2 list the parameters of the distributions of the absolute magnitudes $M_B$ and $M_K$ in $B$ and $K_S$ filters, the linear diameters $A$ (in kpc) and their logarithms, and the logarithms of the nominal total (indicative) masses $M_{25}$ , stellar masses $M_*$ , and hydrogen masses $M_{HI}$ (all masses are in units of the solar mass). The absolute magnitude $M_B$ is taken from the HyperLEDA database. Data for 2301 of the 2MIG galaxies are given there. The absolute $K_S$ magnitude was determined from the apparent magnitude in the 2MIG catalog corrected as pointed out in Section 2. Evolutionary and $K$ corrections were not included. The absorption needed for the corrections was not available for all the galaxies 452
of the 2494 with known radial velocities in HyperLEDA, so the subsample with known $M_K$ was reduced to 2487 galaxies. The distributions of the 2MIG galaxies with respect to $M_B$ and $M_K$ are shown in Figs. 7, a and b. The logarithm of the optical linear diameter A was calculated from the value of the logarithm of the corrected angular diameter from HyperLEDA. A histogram of the distribution with respect to $\log A$ is shown in Fig. 7c. The indicative (total) mass of a galaxy within the limits of the 25 mag/arcsec$^2$ isophote was calculated by usual way
[19] from the rotation velocity $V_{rot}$ and the value of the corrected angular diameter from HyperLEDA. The mass of neutral hydrogen was determined from the radio flux $f$ [19] by relating it to the corrected radio magnitude $m21c=-2.5\log f + 17.40$ from HyperLEDA. A histogram of the distribution on $\log M_{HI}$ is shown in Fig. 7d. For calculating the stellar mass $M_*$ , we assume that $M_*/L_K = 1\cdot M_{\odot}/L_{K,{\odot}}$,  [20]. Here and in the following, on going from absolute magnitudes to luminosities we use the values $M_{K,{\odot}}    =3.28^m$ and $M_{B,{\odot}}=  5.40^m$.
\begin{table}
\begin{center}

\caption{Parameters of the Distributions of the Absolute Characteristics of the Isolated Galaxies}
\begin{tabular}{|l|c|r|r|r|c|} \hline\hline

Sample &  Characteristic& Mean &Min &  Max& $\sigma$ \\
\hline
N = 2301 &$M_B$       &--20.44 & --23.84 &--14.15 &0.98\\
N = 2487 &$M_K$       &--23.71 & --26.62 &--16.76 &1.18\\
N = 1706 &$\log A$     &1.44   & 0.23   &1.99   &0.22\\
N = 1706&   $A$      & 30.9&   1.68 & 97.2& 13.7 \\
N = 978  &$\log M_{25}$   &10.89  & 8.12   &12.51  &0.55\\
N = 2487 &$\log M_*$   & 10.80  & 8.01   &11.96  &0.47\\
N = 977  &$\log M_{HI}$ &  9.70   & 7.15   &11.21  &0.50\\
\hline
\end{tabular}
\end{center}
\end{table}

\begin{figure}
\includegraphics[bb=87 200 508 596,clip,width=.35\textwidth]{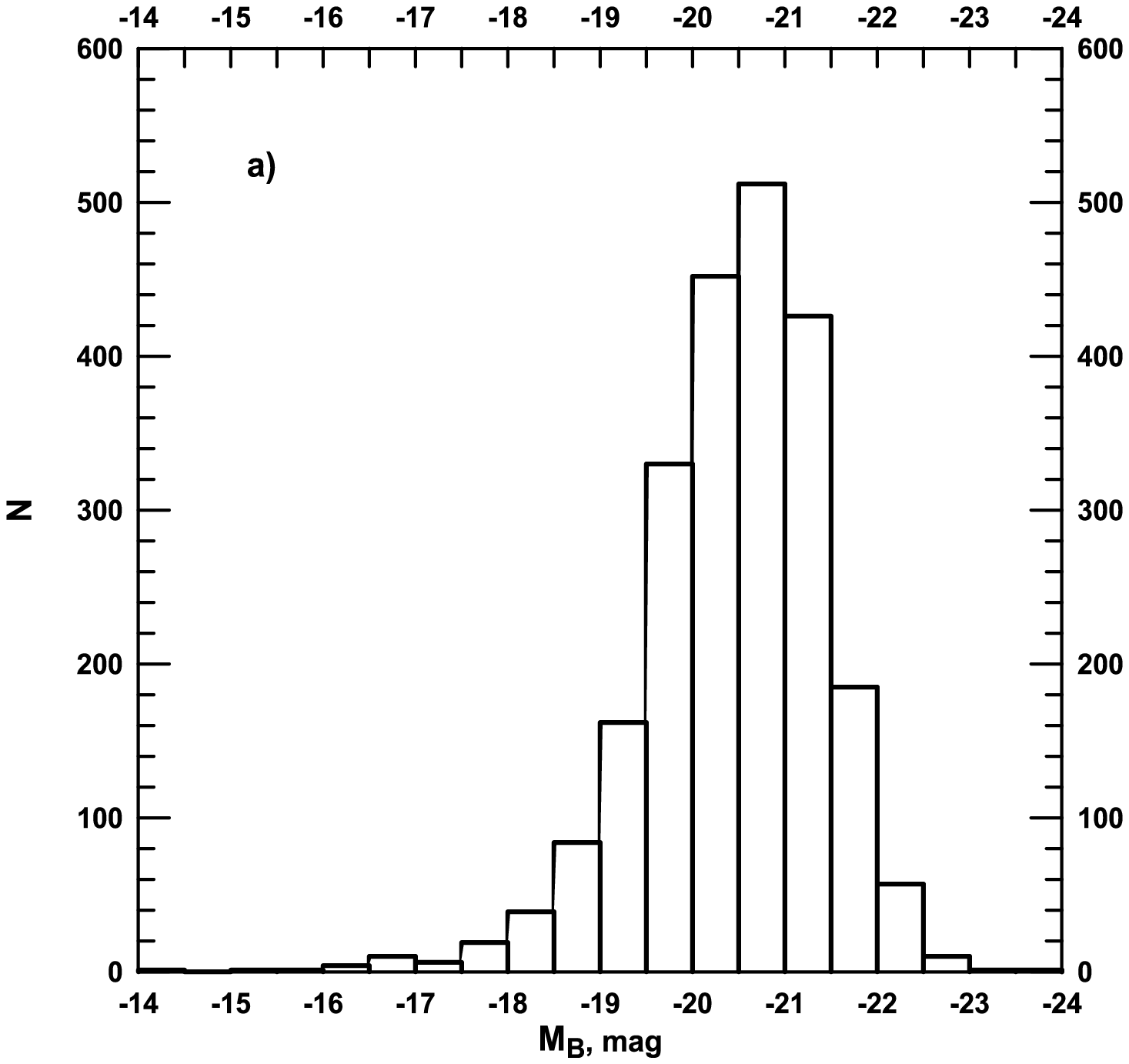}
\includegraphics[bb=87 198 508 596,clip,width=.35\textwidth]{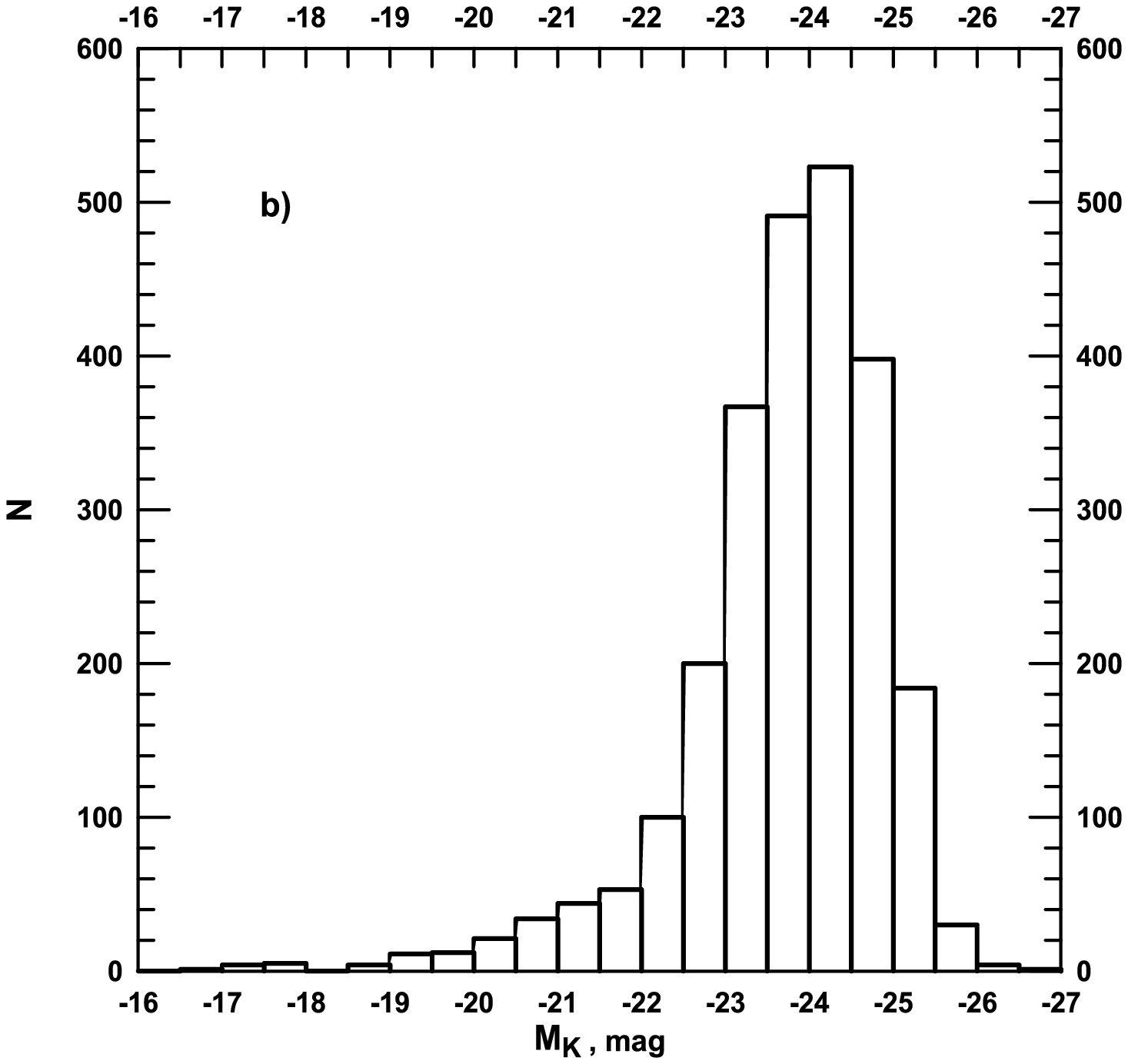}\\ \hfill
\includegraphics[bb=86 127 508 525,clip,width=.35\textwidth]{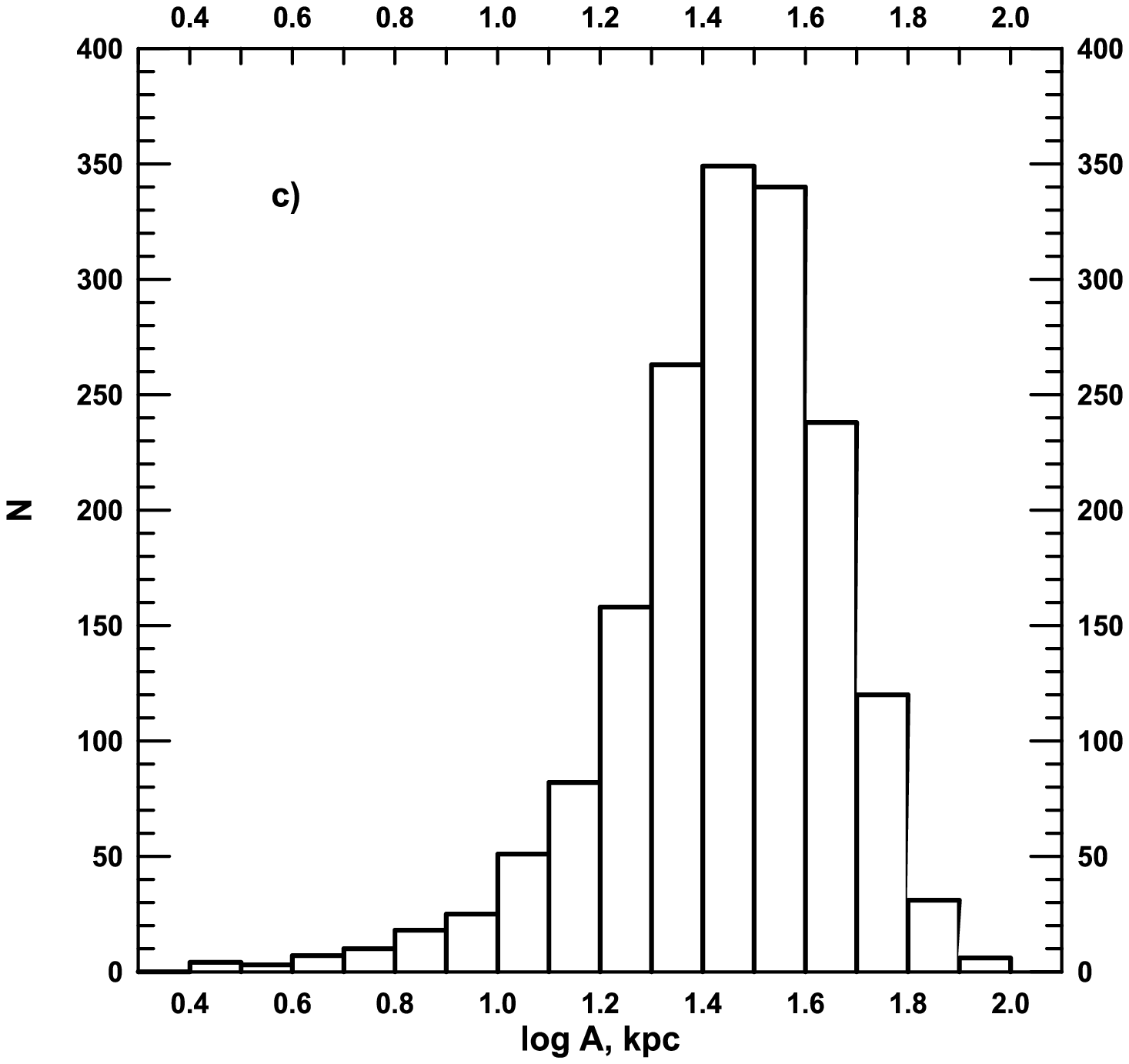}
\includegraphics[bb=86 200 509 596,clip,width=.35\textwidth]{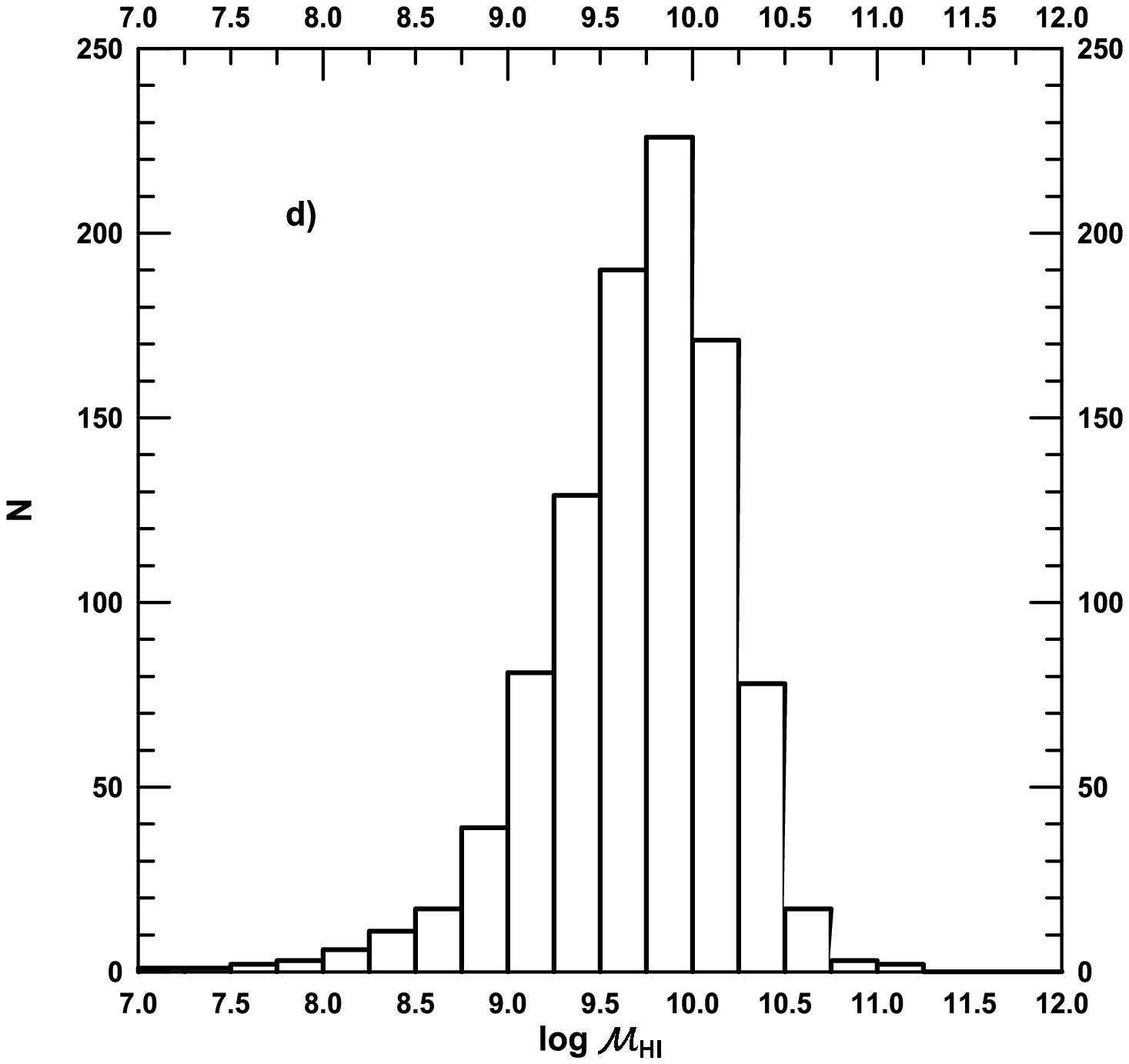} \hfill
\caption{The distribution of the 2MIG galaxies on: (a) absolute $B$ magnitude, (b) absolute $K_S$
 magnitude, (c) the logarithm of the linear diameter, (d) the logarithm of the hydrogen mass.} 
\end{figure}

 From the data in Table 2, it is easy to find that the average ratio of the indicative mass to the stellar mass is 1.2, while the average ratio of the mass of neutral hydrogen to the stellar mass for the 2MIG galaxies is $\sim$1/12.

\bigskip

{\bf \large 5. Some two-dimensional distributions}

\bigskip

 The parameters of the direct and reverse least squares regressions for all these dependences are listed in Table 3. Here $\rho$ is the correlation coefficient and $\sigma$ is the standard deviation from the regression line. The masses $M_{25}$ and $M_*$
 are in units of the Sun's mass, the linear diameter is in kpc, and the rotation velocity is in km/s. The dependences
of the hydrogen mass on the other global characteristics will be given in the next section.

 Table 3 shows that the most highly correlated of the quantities considered here are the absolute $B$ magnitude ($M_B$) and the logarithm of the linear diameter, while the least correlated are $M_B$ and the logarithm of the rotation velocity. Row 7 corresponds to a variant of the Tully-Fisher relation.

\begin{table}
\begin{center}
\caption{Parameters of Linear Least Squares Fits $y  =k \cdot x + b$
for the Global Characteristics of the Isolated Galaxies}
\begin{tabular}{|c|c|c|c|r|c|c|c|} \hline\hline
n& $N$& $y$& $x$& $\rho$& $\sigma$& $k\pm\Delta k$& $b\pm\Delta b$ \\
\hline

1& 1620  &$M_B$ &$\log A$                 &-0.853 & 0.506& -3.86$\pm$0.06   & -14.96$\pm$0.09 \\
 &       &$\log A$&$M_B$                  &       & 0.112& -0.188$\pm$0.003 & -2.42$\pm$0.06 \\
2& 1706  &$M_K$ &$\log A$                 &-0.826 & 0.705& -4.70$\pm$0.08   & -16.84$\pm$0.11 \\
 &       &$\log A$& $M_K$                 &       & 0.124& -0.145$\pm$0.002 & -1.99$\pm$0.06 \\
3&  921  &$M_B$& $\log(M_{25})$           &-0.761 & 0.643& -1.42$\pm$0.04   & -5.1$\pm$0.4   \\
 &       &$\log(M_{25})$& $M_B$           &       & 0.346& -0.409$\pm$0.012  & 2.5$\pm$0.2    \\
4& 978   &$\log(M_{25})$& $\log(M_*)$     &0.814  & 0.319& 0.81$\pm$0.02     & 2.24$\pm$0.20   \\
 &       &$\log(M_*)$ &$\log(M_{25})$     &       & 0.319& 0.82$\pm$0.02     & 1.8$\pm$0.2    \\
5& 978   &$\log(M_{25})$&$\log A$         &0.842  & 0.296& 1.99$\pm$0.04     & 8.07$\pm$0.06   \\
 &       &$\log A$&$\log(M_{25})$         &       & 0.126& 0.357$\pm$0.007   & -2.47$\pm$0.08 \\
6& 978   &$\log A$&$\log(2\cdot V_{rot})$  &0.612  & 0.184& 0.76$\pm$0.03     & -0.49$\pm$0.08 \\
 &       &$\log(2\cdot V_{rot})$&$\log A$  &       & 0.148& 0.49$\pm$0.02     & 1.81$\pm$0.03   \\
7& 1041  &$M_B$& $\log(2\cdot V_{rot})$    &-0.549 & 0.838& -2.96$\pm$0.14   & -13.0$\pm$0.4  \\
 &       &$\log(2\cdot V_{rot})$& $M_B$    &       & 0.155& -0.102$\pm$0.005  & 0.43$\pm$0.10   \\
\hline
\end{tabular}
\end{center}
\end{table}

\begin{table}
\begin{center}
\caption{Parameters of Linear Orthogonal Fits $y=  k\cdot  x + b$ between the Global Characteristics of the Isolated Galaxies}
\begin{tabular}{|c|c|c|c|c|c|} \hline\hline
n &Sample & $\rho$ & $\sigma$& $k\pm\Delta k$& $b\pm\Delta b$\\
\hline
1&2&3&4&5&6\\
\hline
\multicolumn{6}{|c|}{$M_B=y, \,\,\, \log A=x \,\,\, (\xi=1/5)$}\\
\hline
1 & All (1620)         &--0.853& 0.389& --4.452$\pm$0.013& --14.107$\pm$0.019\\
  & Early (648)        &--0.819& 0.394& --4.368$\pm$0.023& --14.138$\pm$0.033\\
  & Intermediate (540) &--0.818& 0.380& --4.470$\pm$0.026& --14.157$\pm$0.039\\
  & Late (432)         &--0.901& 0.378& --4.421$\pm$0.020& --14.190$\pm$0.028\\
\hline
\multicolumn{6}{|c|}{$M_K=y, \,\,\, \log A=x \,\,\, (\xi=1/6)$}\\
\hline
2 &All (1706)         &--0.826 &0.535 &--5.623$\pm$0.015&--15.503$\pm$0.021\\
  &Early (671)        &--0.853 &0.410 &--4.678$\pm$0.018&--17.229$\pm$0.026\\
  &Intermediate (569) &--0.804& 0.447 &--4.663$\pm$0.023&--16.907$\pm$0.034\\
  & Late (466)        &--0.885& 0.532 &--5.899$\pm$0.023&--14.625$\pm$0.033\\
\hline
\multicolumn{6}{|c|}{$M_B=y,\,\,\,\, \log M_{25}=x \,\,\, (\xi=5/9)$}\\
\hline
3 &All (921)          &--0.761 &0.476&--1.880$\pm$0.027& --0.02$\pm$0.30 \\
  & Early (217)       &--0.639 &0.568&--1.609$\pm$0.065& --2.85$\pm$0.70 \\
  & Intermediate (361)&--0.701 &0.473&--1.800$\pm$0.049& --0.95$\pm$0.54 \\
  & Late (343)        &--0.850 &0.392&--2.040$\pm$0.036&   1.67$\pm$0.39 \\
\hline
\multicolumn{6}{|c|}{$\log M_{25}=y, \,\,\,\,\, \log M_* = x \,\,\, (\xi=1)$}\\
\hline
4& All (978)         &0.814 &0.236&1.000$\pm$0.022&  0.27$\pm$0.23\\
 & Early (225)       &0.770 &0.240&1.289$\pm$0.067&--3.00$\pm$0.72\\
 & Intermediate (381)&0.737 &0.215&1.348$\pm$0.059&--3.55$\pm$0.63\\
 & Late (372)        &0.886 &0.200&0.893$\pm$0.024&  1.48$\pm$0.24\\
\hline
\multicolumn{6}{|c|}{$\log M_{25}=y, \,\,\,\,\, \log A   = x \,\,\, (\xi=2/5)$}\\
\hline
5 &All (978)         &0.842 &0.224&2.333$\pm$0.018& 7.580$\pm$0.026\\
  &Early (225)       &0.778 &0.262&2.513$\pm$0.051& 7.385$\pm$0.071 \\
  &Intermediate (381)&0.781 &0.224&2.452$\pm$0.038& 7.407$\pm$0.056 \\
  &Late (372)        &0.899 &0.190&2.163$\pm$0.021& 7.777$\pm$0.029 \\
\hline
\multicolumn{6}{|c|}{$\log A=x, \,\,\,\, \log(2V_{rot})=y \,\,\, (\xi=1)$}\\
\hline
6 &All (978)         &0.612&0.128& 1.42$\pm$0.05& --2.14$\pm$0.13\\
  &Early (225)       &0.518&0.148& 1.17$\pm$0.11& --1.56$\pm$0.27 \\
  &Intermediate (381)&0.515&0.127& 1.29$\pm$0.09& --1.67$\pm$0.22 \\
  &Late (372)        &0.713&0.109& 1.69$\pm$0.08& --2.78$\pm$0.20   \\
\hline
\multicolumn{6}{|c|}{$\log M=B, \,\,\,\,\, \log(2V_{rot})=x \,\,\, (\xi=2/9)$}\\
\hline
7 &All (1041)         &--0.549 &0.606& --6.22$\pm$0.06 & --4.81$\pm$0.14 \\
  & Early (255)       &--0.387 &0.711& --4.07$\pm$0.10 &--10.06$\pm$0.26 \\
  & Intermediate (405)&--0.499 &0.597& --5.88$\pm$0.09 & --5.75$\pm$0.24 \\
  & Late (381)        &--0.676 &0.495& --7.55$\pm$0.09 & --1.62$\pm$0.21 \\
\hline
\end{tabular}
\end{center}
\end{table}
Table 4 lists the parameters of the orthogonal regressions for the pairs of quantities from Table 3, both for
the entire samples (All) and for three subsamples: early galaxies of types E, S0, Sa, and Sab; galaxies of intermediate types Sb and Sbc; and, galaxies of late types from Sc to Im. The types were taken from the 2MIG catalog. The notation is the same as in Table 3. 

The orthogonal regression line coincides with the major diameter of the ellipse obtained assuming a twodimensional gaussian distribution of the points in the plane of the pairs of selected characteristics with and the averages, dispersions, and correlation coefficients obtained for this distribution. The logarithmic maximum likelihood function for this case is $l=-N/2\{1+\ln(2\pi M_1/N)\}$, where $N$ is the number of points in the correlation field of the dependence and $M_1\equiv N\sigma^2_r=\sum(\Delta S_i)^2$.  The measure of the distance between points is $\Delta s^2=(\Delta x/m_x)^2+(\Delta y/m_y)^2$, where $m_x$  and $m_y$ are the axis scales. The parameters of the orthogonal regressions were calculated for the ratios of the scales $\xi\equiv m_x/m_y$,
indicated in Table 4.

In the graphs (Figs. 8--11) of these dependences shown below, we indicate the orthogonal regression line (thick dashed line), the ellipses corresponding to the levels of the maximum likelihood function, and the corresponding 95\% probability, as well as the lines corresponding to the direct and inverse least squares regressions (thin dashed lines), which connect the extreme points of the ellipse along the abscissa and ordinate, respectively. 

The relation between $M_B$ and $\log A$ for the  sample with $N = 1620$ is shown in Fig. 8a. We note the following features of these distributions for the entire sample and subsamples: the correlation coefficient is largest, $\rho=0.90$ , for the sample of late galaxies relative to the samples of early and intermediate galaxies, for which   $\rho=0.82$. The slope of the orthogonal regression depends little on the morphological type. For the orthogonal regression the power-law dependence of $L_B$ on $A$ for the different samples has an exponent in the range $\rho\equiv-0.4k=(1.75\div 1.79)$. This implies, in particular, that the mean surface brightness in the $B$ band falls off slowly with the luminosity  of a galaxy.

 Figure 8b shows the relationship between $M_K$ and $\log A$ for the  sample with $N = 1706$. The correlation coefficient for this relationship was largest for the subsample of late galaxies. The dispersion relative to the orthogonal regression fit increases from the early to the late types; this may be caused by unreliable photometry of the peripheral regions of galaxies without bulges. The slope of the orthogonal regression is largest for the later  
types while the slopes are indistinguishable for the intermediate and early types. The exponent in the dependence of $L_K$ on $A$ lies within the range $\rho=(1.87\div 2.36)$ for the early, intermediate, and late types, with the average ($\rho = 2.35$) indicating a slow growth in the surface brightness with increasing IR luminosity of the galaxies.

\begin{figure}
\begin{center}
\includegraphics[bb=87 127 506 524,clip,width=.4\textwidth]{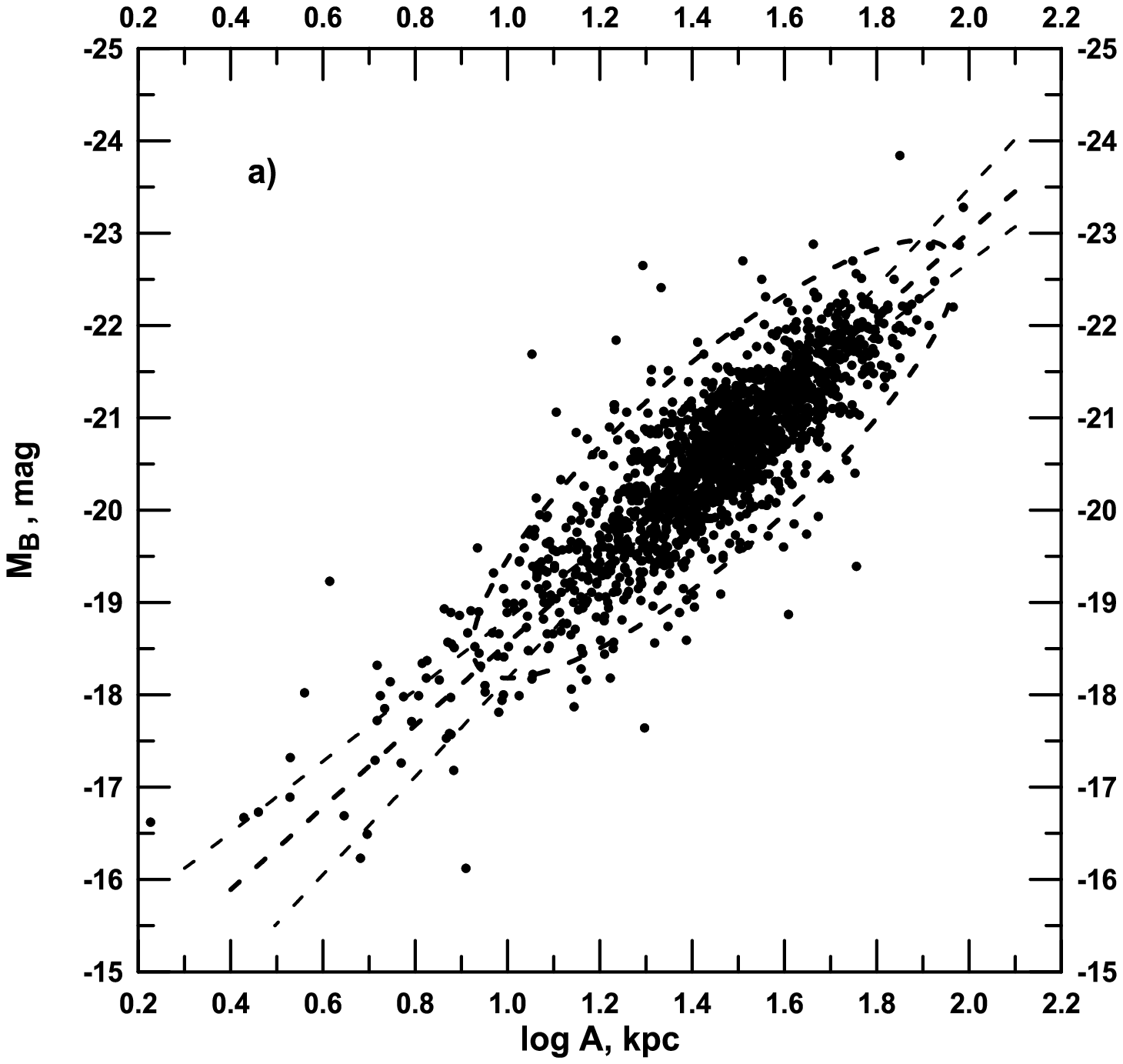}
\includegraphics[bb=87 127 506 524,clip,width=.4\textwidth]{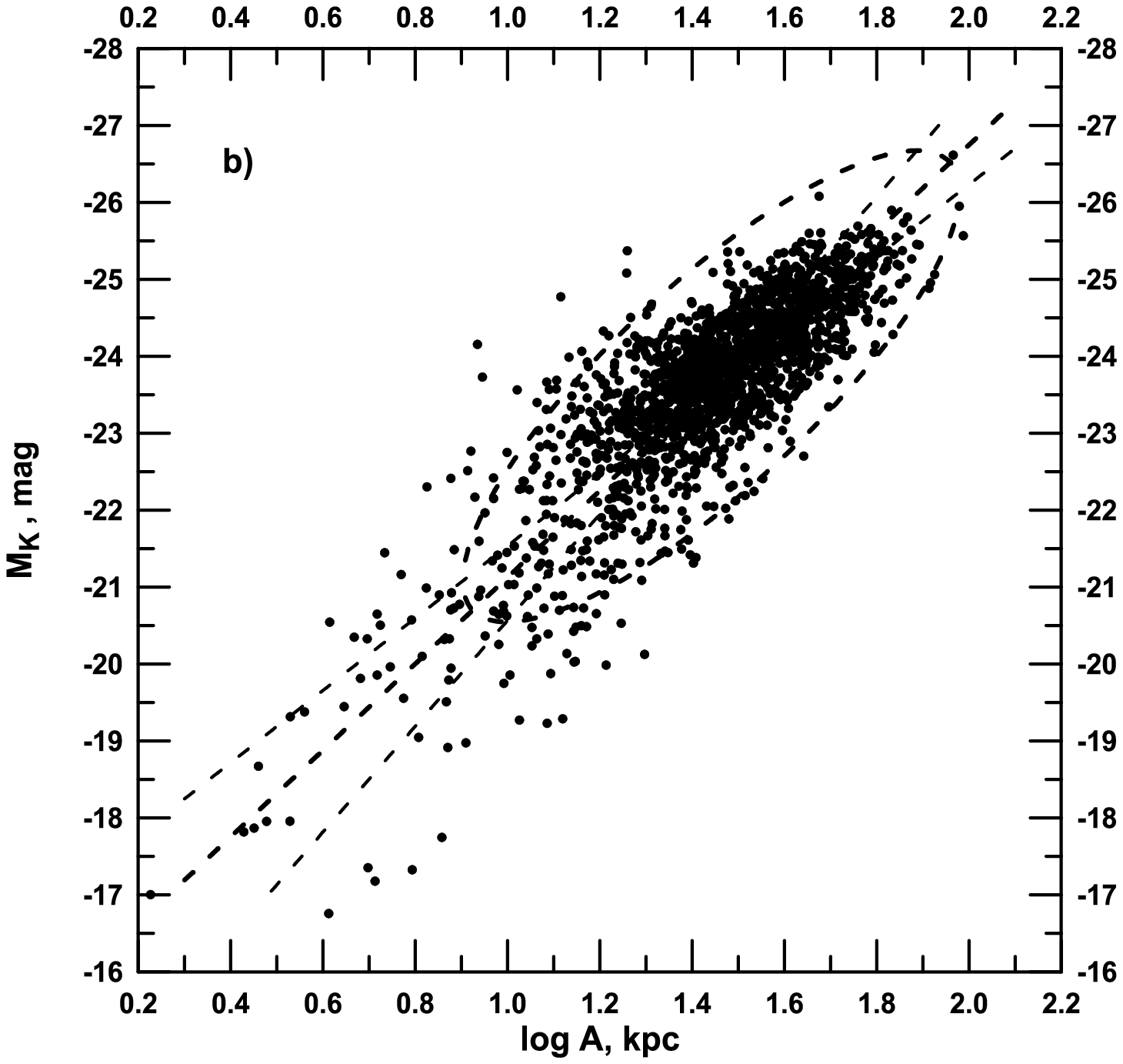}\\ \hfill
\caption{$M_B$ (a) and $M_K$ (b) as functions of $\log A$.} 
\end{center}
\end{figure}

\begin{figure}
\begin{center}
\includegraphics[bb=87 127 506 524,clip,width=.4\textwidth]{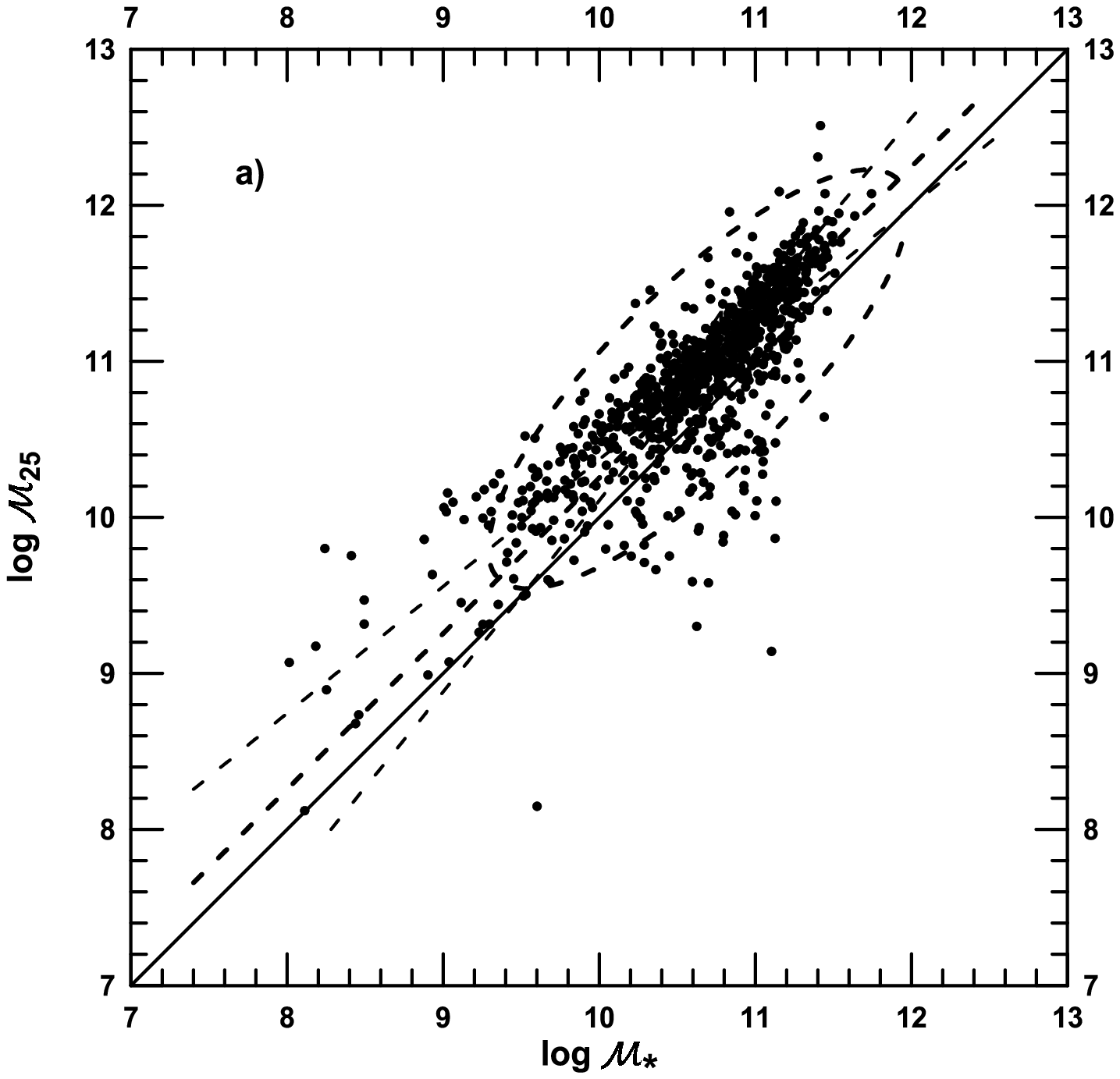}
\includegraphics[bb=87 127 506 524,clip,width=.4\textwidth]{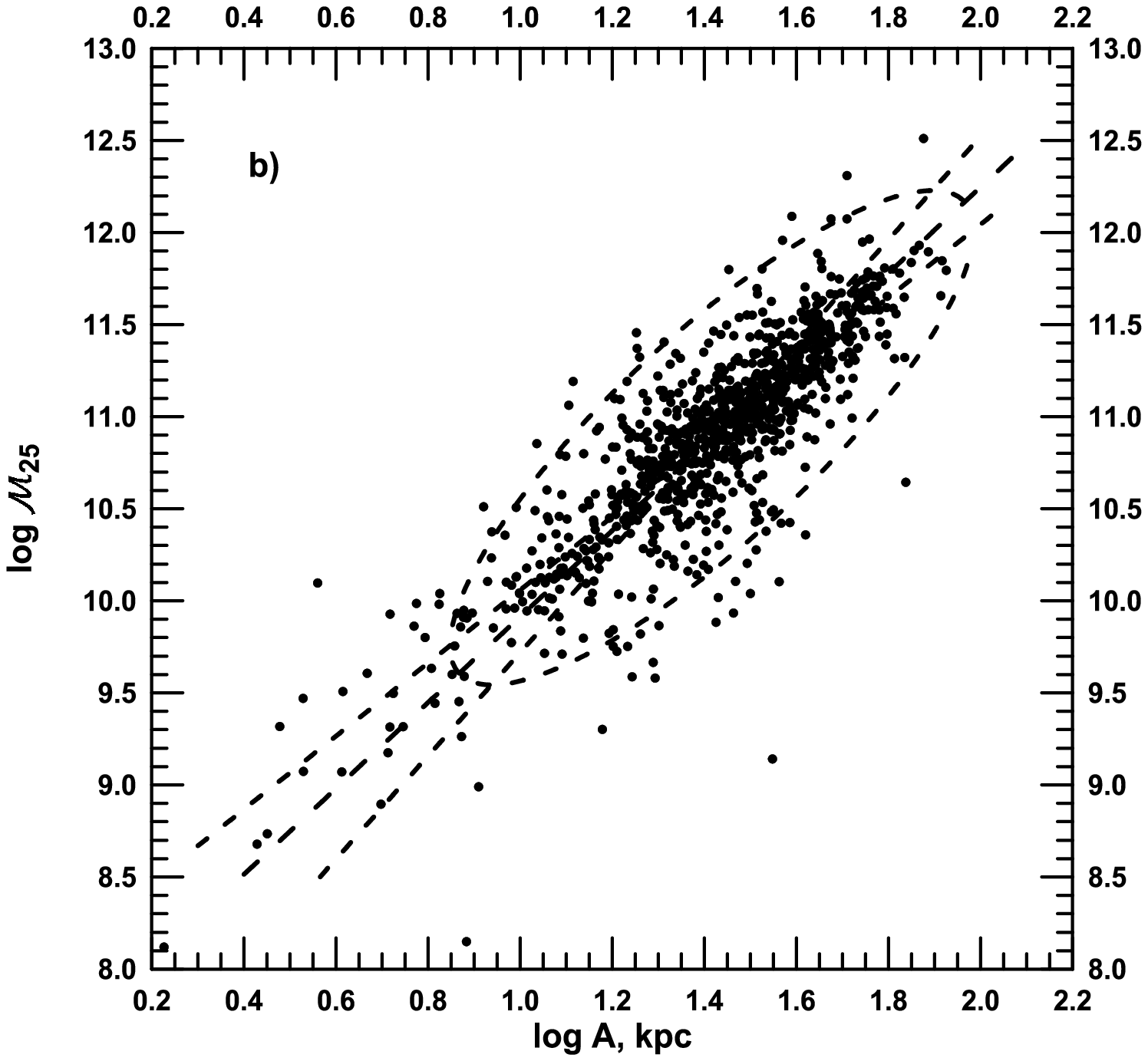}\\ \hfill
\caption{Relationships between (a) $\log M_{25}$ and $\log M_*$, and (b) $\log M_{25}$.}
\end{center}
\end{figure}

The relationship between $\log M_{25}$ and $\log M_*$ for the entire sample with $N = 978$ is shown in Fig. 9a. The solid line indicates the diagonal. For 152 of the 978 galaxies (15.5\%
) the stellar mass $M_*$ was greater than the indicative mass $M_{25}$ (the points lie below the diagonal). Evidently, these are objects with a low surface brightness for which a substantial fraction of the stars lie outside the 25 mag/arcsec$^2$ isophote.

 On the average, for the sample with $N=978$ the indicative mass is 1.8 times greater than the stellar mass. The closest correlation between $M_{25}$ and $M_*$ occurs for the late-type galaxies ($\rho = 0.89$). On the whole, we obtained a simple proportionality between $M_{25}$ and $M_*$ for the 2MIG galaxies $(k= 1.0$), but for the late-type galaxies the slope is smallest ($k = 0.89$) and for the other two subsets the slopes are indistinguishable at $k\approx  1.3$.

 The dependence of $\log M_{25}$ on $\log A$ for the  sample with $N = 978$ is shown in Fig. 9b. According to the data from Table 4, the correlation coefficient between the indicative mass and the linear diameter of the 2MIG galaxies increases from the early to the late types, while the slope of the orthogonal regression decreases toward the late types. The average slope of the orthogonal regression, $k = 2.33\pm0.02$, shows that the average mass surface density increases, while the average volume density falls off with increasing linear size of the galaxies.

\bigskip
{\bf \large 6. Optical and infrared correlators of hydrogen mass}
\bigskip

 We now examine the correlation between the hydrogen mass and the other global characteristics of the isolated galaxies: linear diameter, velocity rotation, $B$ luminosity, and indicative and stellar masses. Table 5 lists the parameters of the linear orthogonal regressions $y=  k\cdot  x + b$ between the logarithm of the hydrogen mass and these characteristics for four samples: All, Early, Intermediate, and Late. The orthogonal regressions were calculated for the scale ratios indicated in Table 5.
\begin{figure}
\begin{center}
\includegraphics[bb=87 127 506 524,clip,width=.4\textwidth]{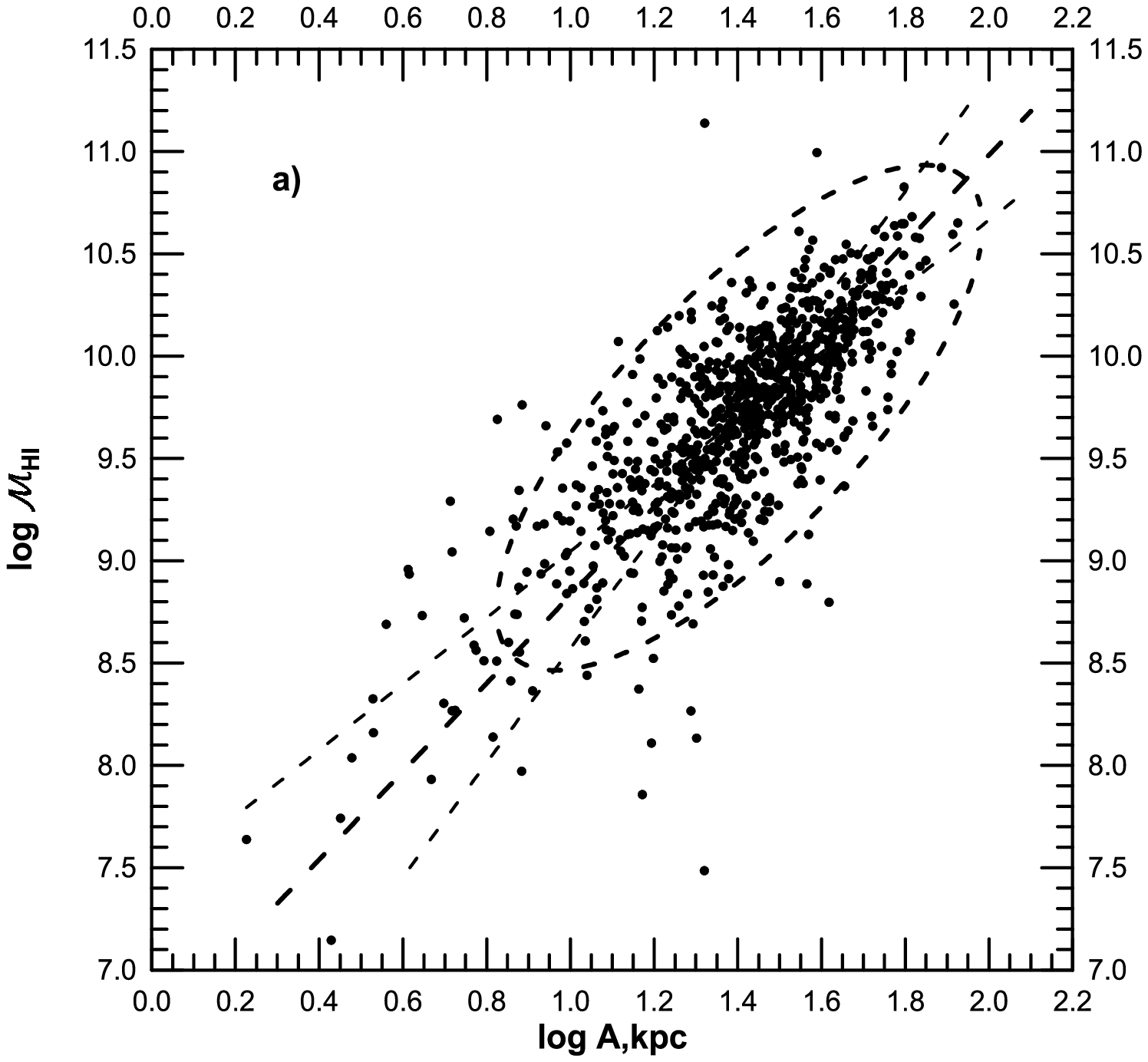}
\includegraphics[bb=87 127 506 524,clip,width=.4\textwidth]{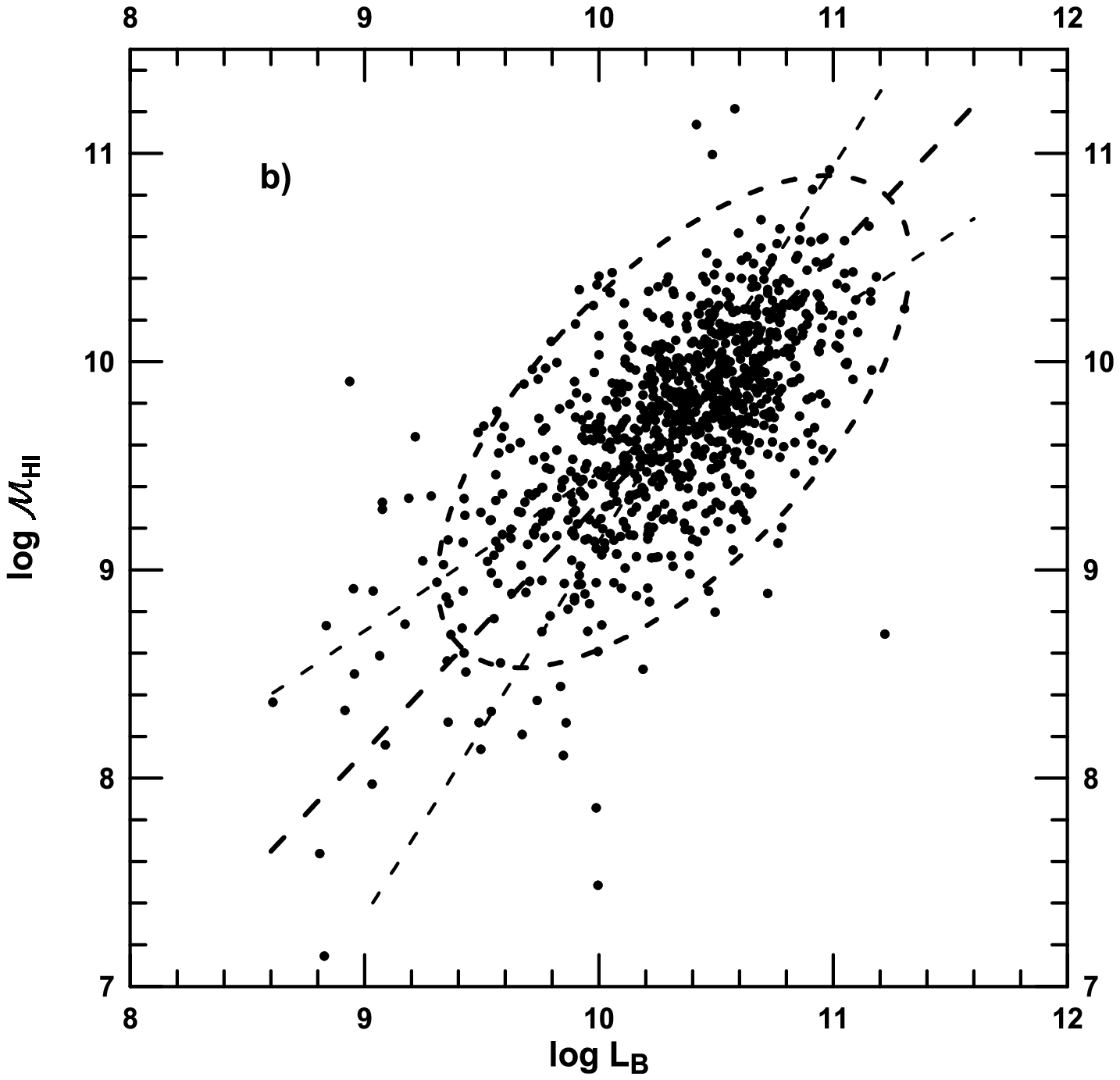}\\ \hfill
\caption{Relationships between (a) $\log M_{HI}$ and $\log A$, and (b) $\log M_{HI}$ and $L_B$.}
\end{center}
\end{figure}

The relationship between $\log M_{HI}$ and $\log A$ is plotted in Fig. 10a. The orthogonal dependence for the three subsamples (of early, intermediate, and late types) has the following properties: (1) the logarithm of the hydrogen mass is most highly correlated with the linear diameter for the galaxies of late types ($\rho = 0.86$) and the least correlated for the early types ($\rho = 0.62$). (2) The slope of the orthogonal regression decreases systematically from early to late types. (3) The samples of galaxies with a dominant disk (intermediate+ late) are characterized by a constant average surface density of the hydrogen mass, independently of the size of the galaxy. We note also that the logarithm of the linear diameter of galaxy turned out to be the best of the six correlators of the hydrogen mass considered here in terms of the value ( $\rho = 0.76$) of the correlation coefficient for the whole sample. 

The relationship between $\log M_{HI}$ and $\log L_B$ plotted in Fig. 10b has the following properties: (1) the best correlation ($\rho = 0.77$) of these characteristics occurs for the late types. (2) The slope of the orthogonal regression decreases systematically from early to late types. (3) The scatter of the galaxies relative to the orthogonal regression decreases systematically on going from early to late types. (4) The average values of the luminosities, as well as of the hydrogen masses, are highest for galaxies of intermediate types. 

The correlation is worst for the relationship between $\log M_{HI}$ and $\log (2V_{rot})  (\rho$ = 0.51). The cause of the weak correlation between the amount of gas and the amplitude of its rotation for galaxies with developed bulges does not appear to be entirely obvious.

The relationship between $\log M_{HI}$ and $\log M_{25}$ is plotted in Fig. 11a. The solid line in Fig. 11a is the diagonal. It can be seen that the hydrogen and dynamic (indicative) masses follow an almost linear relationship. For five galaxies the estimated indicative mass was less than the estimated hydrogen mass. The difference between the average values ($\langle \log M_{25}/M_{\odot}\rangle=  10.86$ and
$\langle\log M_{HI}/M_{\odot}\rangle=  9.70$) corresponds to 7\%
 of the mass of neutral
hydrogen in the mass $M_{25}$. Here the fraction of hydrogen in dwarf galaxies with masses $M_{25}=  10^8 M_{\odot}$ is a factor of three greater than for galaxies with a mass of $10^{12} M_{\odot}$. 

The relationship between $\log M_{HI}$ and $\log M_*$ is shown in Fig. 11b. For 20 of the galaxies the hydrogen

\begin{table}
\begin{center}
\caption{Parameters of Linear Orthogonal Regressions between the
Logarithm of the Hydrogen Mass and other Characteristics}
\begin{tabular}{|c|c|c|c|c|c|} \hline\hline
n &Sample & $\rho$ & $\sigma$& $k\pm\Delta k$& $b\pm\Delta b$\\
\hline
\multicolumn{6}{|c|}{$\log M_{HI}=y \,\,\,  \log A=x \,\,\, (\xi=22/45)$}\\
\hline

1 & All (877)          &0.761& 0.241& 2.15$\pm$0.03& 6.68$\pm$0.04\\
  & Early (211)        &0.617& 0.303& 2.77$\pm$0.11& 5.71$\pm$0.15\\
  & Intermediate (326) &0.728& 0.212& 2.08$\pm$0.05& 6.80$\pm$0.07\\
  & Late (340)         &0.855& 0.201& 1.96$\pm$0.03& 7.00$\pm$0.04\\
\hline
\multicolumn{6}{|c|}{$\log M_{HI}=y \,\,\,  \log L_B= x \,\,\, (\xi=8/9)$}\\
\hline
2 &All (916)          & 0.651 &0.279 & 1.19$\pm$0.04& --2.6$\pm$0.4\\
  &Early (232)        & 0.481 &0.321 & 1.79$\pm$0.16& --8.9$\pm$1.6\\
  &Intermediate (344) & 0.621& 0.262 & 1.15$\pm$0.06& --2.2$\pm$0.6\\
  & Late (340)        & 0.765& 0.240 & 1.07$\pm$0.04& --1.3$\pm$0.4\\
\hline
\multicolumn{6}{|c|}{$\log M_{HI}=y \,\,\,  \log 2V_{rot}= x \,\,\, (\xi=4/9)$}\\
\hline
3 &All (974)          &0.514 &0.321& 2.95$\pm$0.06 &2.34$\pm$0.15\\
  & Early (242)       &0.382 &0.390& 3.02$\pm$0.16 &1.98$\pm$0.40\\
  & Intermediate (362)&0.421 &0.310& 2.71$\pm$0.11 &2.94$\pm$0.27\\
  & Late (370)        &0.679 &0.257& 3.05$\pm$0.07 &2.20$\pm$0.17\\
\hline
\multicolumn{6}{|c|}{$\log M_{HI}=y \,\,\,  \log V_{rot}A = x \,\,\, (\xi=7/9)$}\\
\hline
4& All (876)         &0.732 &0.257&1.33$\pm$0.03&4.91$\pm$0.11\\
 & Early (211)       &0.580 &0.329&1.61$\pm$0.10&3.76$\pm$0.38\\
 & Intermediate (326)&0.679 &0.233&1.26$\pm$0.05&5.17$\pm$0.20\\
 & Late (339)        &0.856 &0.198&1.26$\pm$0.03&5.23$\pm$0.11\\
\hline
\multicolumn{6}{|c|}{$\log M_{HI}=y \,\,\,  \log M_{25}=x \,\,\, (\xi=4/3)$}\\
\hline
5 &All (876)         &0.685 &0.254&0.99$\pm$0.04&--1.03$\pm$0.47\\
  &Early (211)       &0.527 &0.320&1.21$\pm$0.15&--3.65$\pm$1.67 \\
  &Intermediate (326)&0.615 &0.235&0.93$\pm$0.08&--0.39$\pm$0.85 \\
  &Late (339)        &0.829 &0.193&0.93$\pm$0.04&--0.34$\pm$0.47 \\
\hline
\multicolumn{6}{|c|}{$\log M_{HI}=y \,\,\,  \log M_*=x \,\,\, (\xi=4/3)$}\\
\hline
6 &All (977)         &0.643&0.268& 1.00$\pm$0.04& --0.95$\pm$0.48\\
  &Early (242)       &0.538&0.263& 1.66$\pm$0.20& --8.30$\pm$2.13 \\
  &Intermediate (362)&0.606&0.212& 1.34$\pm$0.11& --4.59$\pm$1.19 \\
  &Late (373)        &0.798&0.220& 0.84$\pm$0.04& 0.98$\pm$0.43   \\
\hline
\end{tabular}
\end{center}
\end{table}

\begin{figure}
\begin{center}
\includegraphics[bb=87 127 506 524,clip,width=.4\textwidth]{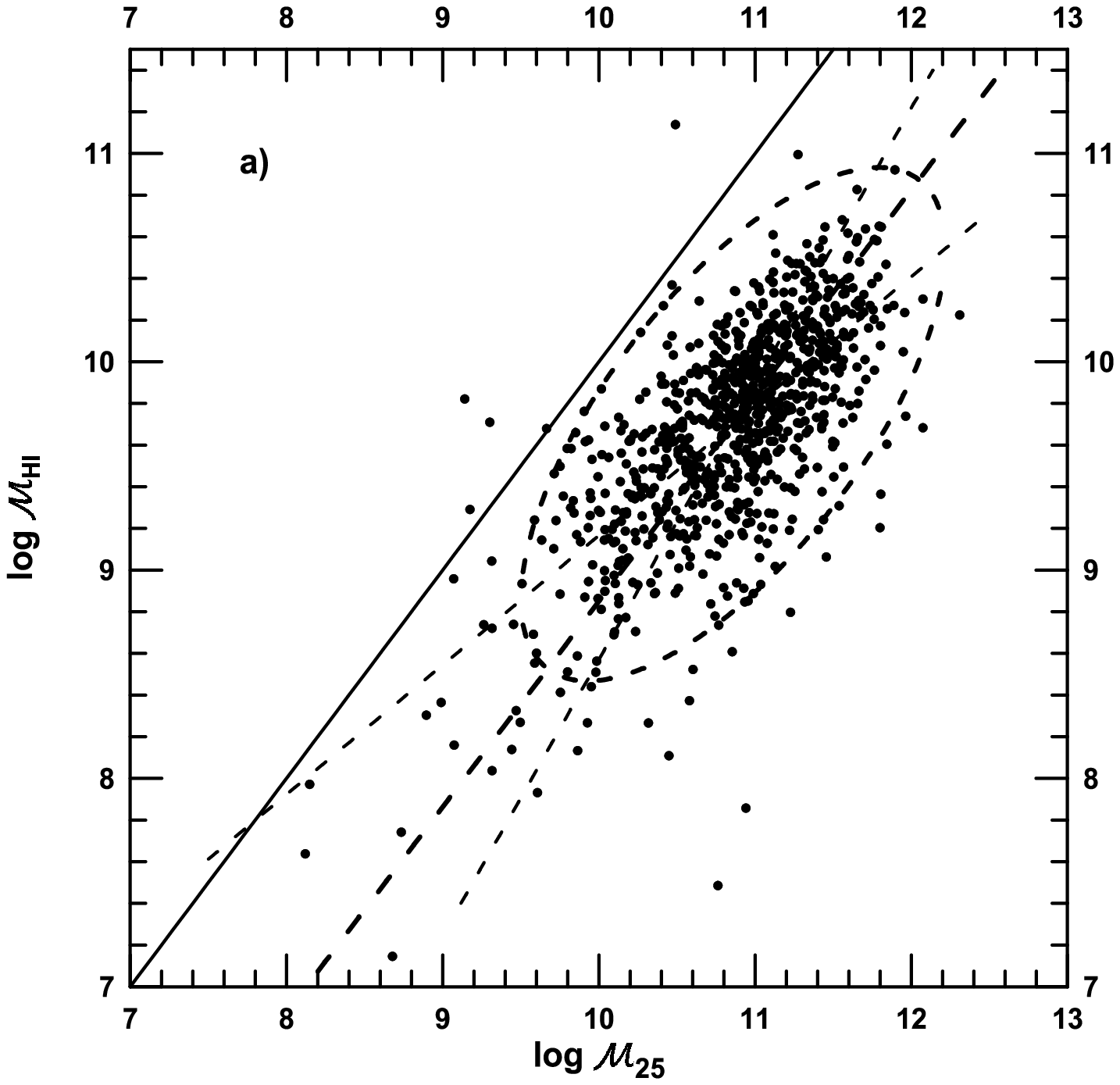}
\includegraphics[bb=87 127 506 524,clip,width=.4\textwidth]{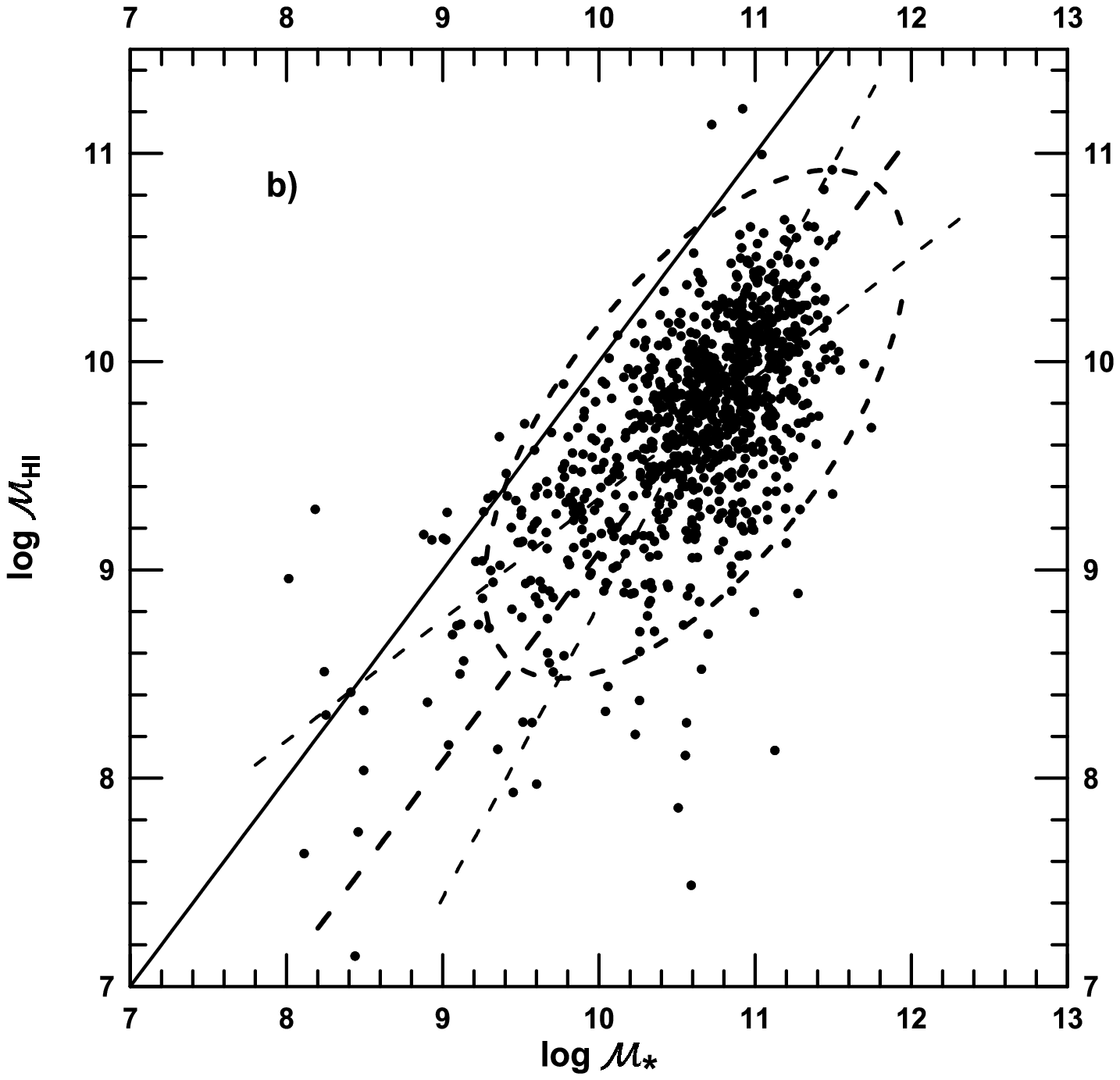}\\ \hfill
\caption{Relationships between (a) $\log M_{HI}$ and $\log M_{25}$, and (b) $\log M_{HI}$ and $\log M_*$.}
\end{center}
\end{figure}

mass exceeds the stellar mass (they lie above the diagonal indicated by the solid line) and for 2 of them the hydrogen mass was also greater than the indicative mass, which suggests a possible error in the data on the hydrogen fluxes in these galaxies. 

The difference between the average values of the logarithms of the masses ($\langle\log M_*/M_{\odot}\rangle=  10.61$ and $\langle\log M_{HI}/M_{\odot}=  9.70$) over the entire sample corresponds to 11\%
 of the fraction of hydrogen in the combined mass
of the stellar and hydrogen component. The slope of the orthogonal regression decreases systematically from the early to the late types, with the change becoming significant at a level of $\sim10\sigma$.

\bigskip
\newpage
{\bf \large 7. Concluding comments}\\ 

\bigskip

 There have been many studies [4,20--24] of the scale relationships between the global parameters of galaxies (luminosity, size, rotation amplitude, neutral hydrogen mass, etc.). Data on the parameters of the scale relationships for different sources often differ substantially. The reasons for these differences may include: (a) selection effects in sampling galaxies as a function of their surroundings (members of clusters, field galaxies); (b) the use of different photometric bands ($B, g, i, K$) or systems of galactic size ($R_{25}, R_{50}, R_{90}$); (c) differences in the morphological composition of the samples and errors in classification that increase with distance to the galaxies; (d) inclusion or exclusion of dwarf galaxies from the nearby volume, where the radial velocity is an unreliable indicator of distance; (e) the use of orthogonal or least squares regressions, the differences between which increase as the correlation coefficient becomes smaller; and, (f) inclusion or neglect of statistical weights proportional to the luminosity or mass of galaxies, as well as other factors. 

In this paper we have established the following characteristics of the scale relationships among the parameters of isolated galaxies from the 2MIG catalog [10]:

(1) The relationship between the $B$ luminosity and the standard isophotal diameter $A\equiv A_{25}$ of the galaxies
0.2 is such that their mean surface brightness decreases slowly with increasing size of a galaxy as $\sim A^{-0.2}$ , regardless of its morphological type. However, in the $K$ band (for $M_*/L_K=  1\cdot M_{\odot}/L_{K,{\odot}})$ the average surface brightness of the stellar mass decreases with the size of a galaxy ($\sim A^{-0.17}$) only for E--Sb morphological types, while it increases as $\sim A^{0.25}$ for Sc- and Sd-galaxies.

(2) The ratio of the indicative mass of a galaxy to its blue luminosity, $M_{25}/L_B$ , increases slightly with
 increasing luminosity as $\sim(L_B)^{0.25}$  for galaxies of both early and late types.

(3) On the average the indicative (dynamic) mass of a galaxy is proportional to its stellar mass, i.e., $M_{25}/M_*\approx$  const, although for galaxies of early types this relation increases slightly with increasing stellar mass (i.e., $K_S$ luminosity). 

(4) The dynamic mass $M_{25}$ of a galaxy within the standard isophote depends on its diameter as  $\sim A^{2.3}$. Note that, in the standard cosmological model the total mass of the dark halo is proportional to the cube of its linear size, i.e., the average density $M_{tot}/A^3$ of the dark halo is assumed to be roughly the same for small light and massive haloes.

(5) The correlation between the rotation amplitude and the diameter for all the galaxies in our sample obeys
$V_{rot}\sim  A^{0.70}$, with a noticeably smaller exponent (0.59) for galaxies of late types without bulges. Note that for dark haloes the expected relationship should be linear, i.e., $V_{rot}/A=$ const. It is interesting that we obtained just this sort of linear relationship in an early study of flat edge-on spiral galaxies  [25]. 

(6) If we exclude galaxies with developed bulges (E, S0, Sa), then the isolated galaxies of types Sb, Sc, and Sd have a constant average surface hydrogen mass density, i.e., $M_{HI}/A^2=$ const, a behavior that has been found previously [20,21].

(7) The ratio of the hydrogen mass to the blue luminosity increases slightly, with $M_{HI}/L_B \sim (L_B)^{0.2}$, for all the morphological types. This tendency does not extend to dwarf galaxies, the relative number of which is only 2\%
 in our sample. 

(8) The relation between  the hydrogen mass and the rotation amplitude is  $M_{HI} \sim (V_{rot})^3$, with a correlation coefficient that increases from 0.38 to 0.68 on going from early to late types.

 (9) On the whole, $M_{HI}/M_{25}$ = const and $M_{HI}/M_*$ = const for the sample of 2MIG galaxies. However, these relations are steeper for galaxies of early types than for the late galaxies. 

Since the isolated galaxies in our sample lie in regions with extremely low density of matter, we can assume that evolutionary merging effects have had little influence on the character of the correlations among the global parameters of the 2MIG galaxies.

 Here we have used information from the HyperLEDA (http://leda.univ-lyon1.fr) and NED  (http://\\nedwww.ipac.caltech.edu), and from the sky surveys DSS--1, DSS--2 (http://archive.eso.org/dss/dss) and SDSS (http://www.sdss.org). This work was partially supported by grants from the Russian Foundation for Basic
Research (RFFI; No. 11--02--90449-Ukr-f-a) and the State Foundation for Basic Research of Ukraine (project F040.2/049).

\bigskip
{\bf \large REFERENCES}
\bigskip

1. D. J. Croton, in: "Galaxies in Isolation: Nature versus Nurture," {\em ASP Conf. Ser.} \/ {\bf 421}, 35 (2010).

 2. S. S. Allam, D. L. Tucker, and B. C. Lee, J. A. Smith, {\em Astron. J.} \/ {\bf 129}, 2062 (2005).

3. H. M. Hernandez-Toledo, J. A. Vazquez-Mata, L. A. Martinez-Vazquez, et al., {\em Astron. J.} \/ {\bf 136}, 2115 (2008).
 
4. M. C. Toribio, J. V. Solanes, R. Giovanelli, et al., {\em Astrophys. J.} \/ {\bf 732}, 93 (2011).

5. F. M. Reda, D. A. Forbes, M. A. Beasley, et al., {\em Mon. Notic. Roy. Astron. Soc.} \/ {\bf 354}, 851 (2004).

 6. V. E. Karachentseva, I. D. Karachentsev, and M. E. Sharina, {\em Astrofizika} \/ {\bf 53}, 513 (2010).

7. J. Sabater, S. Leon, L. Verdes-Montenegro, et al., {\em Astron. Astrophys.} \/ {\bf 486}, 73 (2008).

8. V. E. Karachentseva, {\em Soobshcheniya SAO} No. {\bf 8}, 3 (1973).

 9. F. Zwicky, E. Herzog, P. Wild, et al., Catalogue of galaxies and of clusters of Galaxies, v. I-VI, California Institute of Technology (1961-1968). 

10. V. E. Karachentseva, S. N. Mitronova, O. V. Melnyk, and I. D. Karachentsev, {\em Bull. Spets. Astrofiz. Obs. RAN} \/  {\bf 65}, 1 (2010) (ftp://cdsarc.u-strasb.fr/ pub/cats/YII/257).

11. M. F. Skrutskie, S. E. Schneider, R. Steining et al, In: {\em The Impact of Large Scale Near-IR Sky Surveys}, ed. F. Garzon et al. (Netherlands: Kluwer), {\em ASSL} \/ {\bf 210}, 25 (1997).

 12. T. N. Jarrett, T.Chester, R.M.Cutri, et al., {\em Astron. J.} \/ {\bf 119}, 2498 (2000).

 13. D. J. Schlegel, D. P. Finkbeier, and M. Davis, {\em Astrophys. J.} \/ {\bf 500}, 525 (1998).

14. I. D. Karachentsev and D. I. Makarov, {\em Astron. J.} \/ {\bf 111}, 794 (1996).

15. L. Verdes-Montenegro, J. Sulentic, U. Lisenfeld et al., {\em Astron. Astrophys.} \/ {\bf 436}, 443 (2005).

16. M. C. Toribio, J. M. Solanes, R. Giovanelli, and M. P. Haynes, {\em Astrophys. J.} \/ {\bf 732}, 92 (2011).

 17. M. Schmidt, {\em Astrophys. J.} \/ {\bf 151}, 393 (1968).

18. V. E. Karachentseva, {\em Astron. zh.} \/ {\bf 57}, 1153 (1980).

19. I. D. Karachentsev, V. E. Karachentseva, W. K. Huchtmeier, and D. I. Makarov, {\em Astron. J.} \/ {\bf 127}, 2031 (2004).

20. I. D. Karachentsev and A. M. Kut'kin, {\em Astron. Lett.} \/ {\bf 31}, 299 (2005).

 21. M. S. Roberts and M. P. Haynes, {\em Ann. Rev.  Astron. Astrophys.} \/ {\bf 32}, 115 (1994).

22. M. P. Haynes and R. Giovanelli, {\em Astron. J.} \/ {\bf 89}, 758 (1984).

23. E. E. Salpeter and G.L.Hoffman, {\em Astrophys. J.} \/ {\bf 465}, 595 (1996).

24. M. J. Disney, J. D. Romano, D. A. Garcia-Appadoo, et al., {\em Nature} \/ {\bf 455}, 1082 (2008).

 25. I. D. Karachentsev, V. E. Karachentseva, and Yu. N. Kudrya, {\em Pis'ma v Astron. zh.}\/ {\bf 25}, 3 (1999).
\end{document}